\documentclass[12pt,twoside,titlepage]{article}
\usepackage[blocks]{authblk}
\usepackage[utf8]{inputenc}
\usepackage{amsmath,amssymb,amscd,amsfonts, amsthm,caption,color,epsfig,
fancyhdr,graphicx,latexsym,mathrsfs,multicol,setspace,wrapfig}
\usepackage{psfrag}
\usepackage{a4wide}
\usepackage{cancel,soul,url}%
\usepackage[
  bookmarksopen,
  colorlinks,
  linkcolor = black,
  urlcolor  = blue,
  citecolor = blue,
  menucolor = black
]{hyperref}

\onehalfspace
 \definecolor{qqffff}{rgb}{0,1,1}
\definecolor{ttffff}{rgb}{0.2,1,1}

\begin{document}
\title{\bf Estimating, Monitoring, and Forecasting the Covid-19 Epidemics: A Spatio-Temporal Approach Applied to NYC Data}
\author[a]{Vinicius V. L. Albani}
\author[b]{Roberto M. Velho} 
\author[c]{Jorge P. Zubelli}

\affil[a]{Department of Mathematics, Federal University of Santa Catarina, Campus Trindade, 88.040-900 Florianopolis-SC, Brazil, \href{mailto:v.albani@ufsc.br}{\tt v.albani@ufsc.br}}
\affil[b]{Institute of Informatics, Federal University of Rio Grande do Sul, PO Box 15064,
Av. Bento Gonçalves, 9500, 91501-970 Porto Alegre-RS, Brazil, \href{mailto:roberto.velho@gmail.com}{\tt roberto.velho@gmail.com}}
\affil[c]{Mathematics Department, Khalifa University, PO Box 127788, Abu Dhabi, UAE, \href{mailto:jorge.zubelli@ku.ac.ae }{\tt jorge.zubelli@ku.ac.ae}}

\date{\today}

\maketitle

\begin{abstract}
We propose an SEIR-type meta-population model to simulate and monitor the Covid-19 epidemic evolution. The basic model consists of seven compartments, namely susceptible (S), exposed (E), three infective classes, recovered (R), and deceased (D). We define these compartments for $n$ age and gender groups in $m$ different spatial locations. So, the resulting model has, for each age group, gender, and place, all epidemiological classes. The mixing between them is accomplished by means of time-dependent infection rate matrices. The model is calibrated with the curve of daily new infections in New York City and its boroughs, including census data, and the proportions of infections, hospitalizations, and deaths for each age range. We end up with a model that matches the reported curves and predicts accurately infection information for different places and age classes.
\end{abstract}

\tableofcontents
\section{Introduction}\label{sec:introduction}

During December 2019, in Wuhan, China, many cases of severe acute respiratory syndrome, caused by an unknown virus, were registered. Since then, the virus was named Sars-Cov-2 and the corresponding disease was called by the acronym Covid-19, which means Coronavirus Disease 2019. 
On 11-Mar-2020, a pandemic was declared by the  World Health Organization (WHO). Then, by 08-Oct-2020 Sars-Cov-2 has infected more than 36 million individuals and has caused more than one million deaths around the world.

One clear message from the above history of the pandemics so far is that the study and management of the crisis calls for the use of spatial-temporal epidemiological models and to their appropriate calibration by means of mathematical tools from numerical analysis and regularization theory \cite{ern,somersalo}. This is the first goal of the present article.

It is now well-documented that the chance of developing the more severe form of the disease increases dramatically with age \cite{verity2020,wu2020,wu2020b,who2020,covid2020severe}. In addition, the infection shows to be more frequent in older people \cite{covid2020severe}. This indicates that age plays an important role in possible containment and mitigation measures. Gender also seems to play a major role on Covid-19 outcomes, as documented in \cite{gender1,gender2}. 
Incorporating age and gender structure in the model is the second goal of this article.

The absence of a universally available vaccine and of an effective treatment, by the time this article is written, limits considerably the number of possible actions to control the spread of the disease and the subsequent volume of hospitalizations and deaths. 
Thus, only containment or mitigation policies, like quarantine or lockdown may be applied. However, such measures have been causing an unprecedented impact on the economy and labor market, leading to massive unemployment and recession \cite{imfTrack}. The International Monetary Fund has revised its forecast in April 2020 and it predicted a 4.9\% drop in global output in 2020 \cite{wef}. 

Quantifying and tracking the disease spread in different places and age ranges, as well as its impact on the health system, is useful to decide if lockdown measures can be relaxed, allowing then the return of economic activities gradually. This is the third goal of this work.

Moreover, an accurate forecasting of the number of regular hospital and intensive care unit (ICU) beds allows a better use of public resources, bringing economic relief.

Susceptible-Exposed-Infected-Recovered (SEIR) and Susceptible-Infected-Recovered (SIR) models have been used to describe different disease outbreaks dynamics since the seminal work \cite{KerMack1927}. See also \cite{keeling2008} in a textbook format. Recently, several authors \cite{tsakris2020,somersalo2020,calvetti2020,dehning2020,gatto2020,wu2020,wu2020c} have applied SIR- and SEIR-type models to describe the Covid-19 epidemic, including different features, like geographical information and time-dependent transmission parameters.

In this article, we also propose a versatile SEIR-type model applied to Covid-19 epidemic dynamics. Our model takes into consideration different levels of disease severity, its impact on age ranges, and the distribution of the population in different locations. Following \cite{verity2020}, individuals in a severe state of this disease are accounted in our model as hospitalized, while those critically ill are considered in an intensive care unit (ICU). 
The interaction between classes of infected and susceptible individuals from different age-ranges, genders, and places is defined by time-dependent transmission matrices. Such matrices, if appropriately calibrated with up-to-date data on daily new infections, can be used to reconstruct the status of the disease spread and they allow us to verify the impact of containment measures. 

Concerning vaccines, the flexible general form of our model can be used to design vaccination strategies that account for age, gender, and spatial distribution of susceptible population. From the sanitary point of view, such designed strategies may break the transmission chain of the disease, while optimizing the costs of immunization of the population on the financial side. 

The dependency of disease severity on age-range and gender is translated into the model through transmission rates, as well as the rates of recovery, hospitalization, ICU admission, and death. The values of those rates are based on publicly available datasets and recent studies that analysed, among other characteristics, the relationship between Covid-19 severity and age \cite{wu2020c,covid2020severe,verity2020,NYCgithub} or gender \cite{gender1,gender2}. Other parameters, like mean incubation time and case fatality rate outside ICU, were obtained from \cite{Incubation1,grasselli2020,guan2020,huang2020,wu2020b,who2020}.

As mentioned above, our proposed model also accounts for spatial information. Policy makers can then identify clusters of uncontained disease spread in real time, isolate them, and later verify if the chosen imposed restrictions measures were effective. Moreover, the model can be used to detect which regions should be reopened first, thus reducing the impact a lockdown creates on economy. Once an effective vaccine is available, the model could be used to target regions where there are clusters of infected individuals. This approach would then be used to create immunization belts around such regions. 
Moreover, the proposed model is able to forecast future spatial- and age-distributed clusters of infected individuals and bring information to design contention or immunization measures.

We end up with a model that is useful to track and forecast epidemics caused by new emerging pathogens, including Sars-Cov-2, in different geographical scales, gender, times, and age classes. The model is easy to implement, since the set of ordinary differential equations (ODE) can be solved by general-purpose ODE solvers. Furthermore, the model is simple to calibrate, with, for example, gradient-based optimizers or Bayesian inference algorithms. 

We obtain the geographical distribution of the disease dynamics considering the five NYC boroughs (Manhattan, Bronx, Brooklyn, Queens, and Staten Island) using the census data \cite{BaruchColege}, the curve of daily new infections \cite{NYCgithub}, and the corresponding proportions of hospitalizations and deaths depending on age classes, by borough.

The estimation of the parameters is performed by minimizing a log-posterior density with a gradient-descent technique. Bootstrap sampling is used to test parameters sensitivity as well as to provide 90\% confidence intervals \cite{chowell2017}.

\paragraph*{Main Findings} 
After smoothing out the daily curves through $\text{7-}$days moving averages, we estimate the model parameters. The predicted curve by the model for daily new infections has good adherence to the averaged data curve. 
Furthermore, the predictions of hospitalizations and deaths match well
the reported values based on NYC data and its boroughs. 

We observe a dramatic change in the pattern of disease transmission on 19-Mar-2020, identifying the effectiveness of containment measures imposed a week earlier, when a state of emergency was declared and people were asked to stay home. We can also observe a considerable drop of the time-dependent transmission coefficient and the time-dependent basic reproduction rate (obtained via the next-generation matrix technique \cite{Diekmann1990}). We also noticed this phenomenon in the dynamics of the time-dependent transmission coefficients associated to the NYC boroughs.

When analysing the datasets, the patterns of daily hospitalizations and deaths changed consistently between the end of February and the end of August, especially the rate of hospitalization, which has decreased systematically since the end of March. To account for this feature, we allow the model rates of hospitalization and death to be time-dependent.
This produces model predictions adherent to the datasets. 

Short-term forecasts, with calibrated parameters, were also tested in two different situations, namely, during the transmission regime change and after the spread containment. In both cases, the model predicted accurately the observed scenarios.

Moreover, different reopening scenarios were simulated, considering the impacts of reopening the entire NYC, the Staten Island only, or schools only. In all such cases, unless strict social distance measures were kept, model predictions indicate new infection waves that affect the population of the entire city (Figs.~\ref{fig:forecast3}-\ref{fig:liftB}). Such findings are corroborated by recent news, with new infection waves identified in Europe, New Zealand, and China \cite{bbc2020,ft2020,euronews}, as well as the reports of Covid-19 spread amongst youth population on an overnight camp in Georgia (United States) and at schools in Israel \cite{CDC2020,NYT}.

\section{Materials and Methods}
This section presents the epidemiological model and the procedure to estimate the model parameters from the reported data.

\subsection{The Epidemiological Model}

The SEIR-type model considered here accounts for disease severity, age ranges, gender, and geographic distribution of some pre-defined group or population. For simplicity, we postpone the inclusion of gender and (spatial) location dependence to the end of the present section. A number $n$ of age ranges is assumed, each one represented by the superscript $i = 1,\ldots,n$ and distributed in seven epidemic compartments: susceptible ($S^i$), exposed but not yet infective ($E^i$), infective in mild conditions ($I^i_M$), infective in severe condition or hospitalized ($I^i_H$), infective in critical condition or in an intensive care unit (ICU), denoted by $I_I^i$, recovered ($R^i$){,} and deceased ($D^i$). Following \cite{verity2020}, we assume the following forms are synonyms: in severe condition and hospitalized. The same applies for the forms in critical conditions and in ICU. Each individual in the first two infective compartments, mildly infective ($M$) and hospitalized ($H$), can recover, die, or develop a more severe disease outcome. Those ones in ICU can only recover or die. 

\begin{figure}[!htb]
  \centering
      \includegraphics[width=0.7\textwidth]{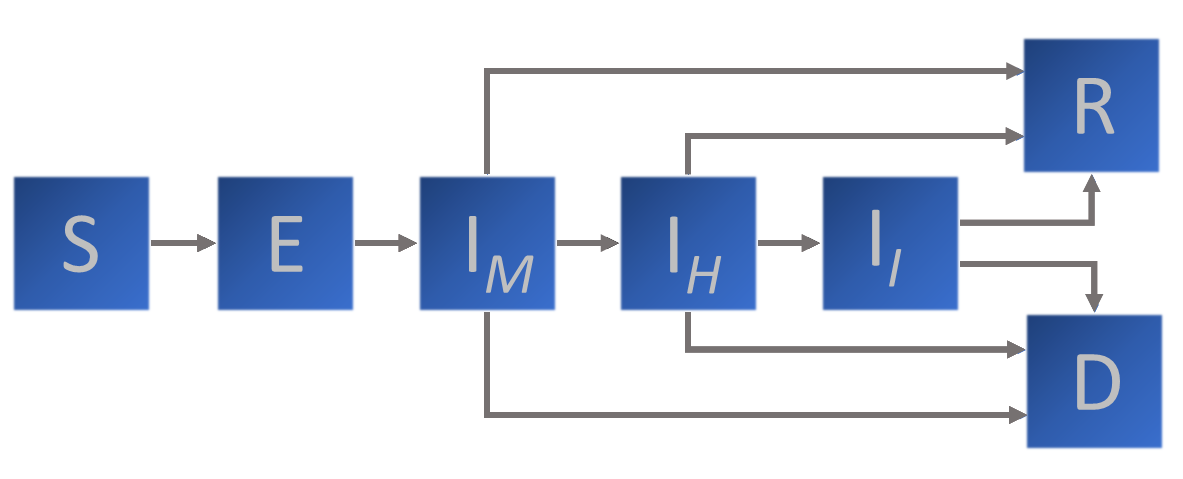}\hfill
     \caption{Schematic representation of the epidemiological model in
     the system of Eqs.~(\ref{eq:modela})-(\ref{eq:model}).}
  \label{fig:SEIR}
\end{figure}

To describe the model, we introduce the following notation: Define the vector
\begin{equation*}
  \textbf{S} = [S^1,\ldots,S^n]^T,
\end{equation*} where the superscript $T$ denotes the transposed vector,
and similarly for $\textbf{E}$, $\textbf{I}_M$, $\textbf{I}_H$, $\textbf{I}_I$, $\textbf{R}$, and $\textbf{D}$. Define also the tensor product:
$$\textbf{x}:\textbf{y}= [x_1y_1,x_2y_2,\ldots,x_ny_n]^T.$$
Then, the epidemiological model can be written as:
\begin{align}
 \dfrac{d\textbf{S}}{dt} & = - S\left(\beta_M \textbf{I}_M + \beta_H \textbf{I}_H + \beta_I \textbf{I}_I\right)\mbox{,} \label{eq:modela}\\
 \dfrac{d\textbf{E}}{dt} &= S\left(\beta_M \textbf{I}_M + \beta_H \textbf{I}_H + \beta_I \textbf{I}_I\right) - \sigma \textbf{E}\mbox{,}\\
  \dfrac{d\textbf{I}_M}{dt} &=  \sigma \textbf{E} - \left(\nu_M + \mu_M + \gamma_M\right):\textbf{I}_M\mbox{,}\\
  \dfrac{d\textbf{I}_H}{dt} &=  \gamma_M:\textbf{I}_M -\left(\nu_H + \mu_H + \gamma_H\right):\textbf{I}_H\mbox{,}\\
  \dfrac{d\textbf{I}_I}{dt} &=  \gamma_H:\textbf{I}_H -\left(\nu_I + \mu_I\right):\textbf{I}_I\mbox{,}\\
  \dfrac{d\textbf{R}}{dt} &=  \nu_M:\textbf{I}_M + \nu_H:\textbf{I}_H + \nu_I:\textbf{I}_I\mbox{,}\\
  \dfrac{d\textbf{D}}{dt} &= \mu_M:\textbf{I}_M + \mu_H:\textbf{I}_H + \mu_I:\textbf{I}_I\mbox{.}
  \label{eq:model}
\end{align}
The schematic representation of the model in Eqs.~(\ref{eq:modela})-(\ref{eq:model}) can be seen in can be seen in
Fig.~\ref{fig:SEIR}.

The matrices $\beta_M$, $\beta_H$, and $\beta_I$ contain the transmission parameters for the infective individuals in the mild, hospitalized, and ICU classes, respectively. Such parameters are time-dependent and, if well calibrated, may be used to address the effectiveness of contention measures, to verify changes in the transmission pattern, or yet to track the impact of suspending lockdown. It is worth mentioning that, depending on the information available, simplifying assumptions on the structure of such matrices must be made. In our study, $\beta_M$, $\beta_H$, and $\beta_I$ assume the following form:
\begin{equation}
 \beta_M = \beta(t)\,\text{B}, ~\beta_H = a\beta_M,~\mbox{and}~\beta_I = b\beta_M,
 \label{eq:betaM}
\end{equation}
where the $n\times n$ matrix $\text{B}$ is of the form:

\begin{equation}
\text{B} = \left[
\begin{array}{ccccc}
 a_1 & a_1b_1 & a_1b_2 & \cdots & a_1b_{n-1}\\
 a_1b_1 & a_2 & a_2b_1 & \cdots & a_2b_{n-2}\\
 \vdots & \ddots & \ddots & \cdots & \vdots\\
 \vdots & \ddots & \ddots & \ddots & \vdots\\
 a_1b_{n-1} & \cdots & a_{n-2}b_2 & a_{n-1}b_1 & a_n
\end{array}
\right],
\label{eq:matrixB}
\end{equation}
{$\beta(t)$} is a time-dependent scalar parameter that controls the epidemic dynamics, and the parameters $a$ and $b$ are scale factors between $0$ and $1$. The matrix $\text{B}$ depends on $2n-1$ parameters, namely, the vectors $\textbf{a} = [a_1,\ldots,a_n]^T$, which contains the observed rates of infections in the $n$ age ranges reflecting the observed heterogeneity of the infectiousness of Covid-19 into the model, and $\textbf{b} = [b_1,\ldots,b_{n-1}]^T$, which addresses the mixing between different age ranges and shall be estimated from the available data. The $n$-dimensional vectors $\nu_M$, $\nu_H$, and $\nu_I$ contain the recovery rate for each age range in the infective classes $M$, $H$, and $I$, respectively. Similarly, $\mu_M$, $\mu_H$, and $\mu_I$ contain the mortality rate for mild, hospitalized, and in ICU infective individuals. The mean time of incubation is $1/\sigma$. The rates $\gamma_M$ and $\gamma_H$ stand for the hospitalization and ICU admission, respectively.

The rates of recovery, mortality, hospitalization, and ICU admission are inversely proportional to the corresponding mean times of disease evolution and directly proportional to the  probabilities of moving on to other compartments. 
All the quantities defining such rates are based on the results of references \cite{covid2020severe,grasselli2020,guan2020,huang2020,NYCgithub,Incubation1,verity2020,who2020,wu2020b}.

The time-dependent transmission parameters $\beta_M$, $\beta_H$, and $\beta_I$, as well as the initial number of infective cases, are unknowns and shall be estimated from the recorded data of daily new infections. Available census data is used to determine the population size and the proportions of susceptible population on each age range.

Moreover, whenever the data from daily reports of new cases (infections, hospitalizations, and deaths) include different age ranges, the model can be generalized to incorporate such information. In this case, the entries of $\beta_M$ are as follows:

\begin{equation*}
[\beta_M]_{jj} = \beta^j(t)B_{jj}, ~[\beta_M]_{ij} = \frac{\text{B}_{ij}}{2}\left(\beta^i(t) + \beta^j(t)\right),~i\not=j,
\end{equation*}
with $\beta^j(t)$, $j=1,\ldots,n$, time-dependent scalar coefficients, and $\text{B}_{ij}$ the entries of the matrix $\text{B}$ defined above. The other transmission parameters are 
{still of the form}
$\beta_H = a\beta_M$ and $\beta_I = b\beta_M$.

Since for the NYC datasets only the accumulated numbers of infections, hospitalizations and deaths are age structured, we assume that the daily reported cases are not age structured. To introduce more realistic death and hospitalization rates, we adjust $\mu_I$ and $\gamma_M$ by appropriate delayed ratios from daily reports. More precisely, if $\overline{\gamma}_M$ and $\overline{\mu}_I$ represent the mean rates of hospitalization and death, respectively, for each age range $i=1,\ldots,n$, the constant rates $\gamma^i_M$ and $\mu^i_I$ are replaced, respectively, by

$$
\gamma^i_M\frac{\tilde{I}_H(t)}{\overline{\gamma}_M \tilde{I}_M(t-\tau_M)}\quad\mbox{and}\quad
\mu^i_I\frac{\tilde{D}(t)}{\overline{\mu}_I \tilde{I}_H(t-\tau_D)},
$$
where $\tilde{I}_M$, $\tilde{I}_H$, and $\tilde{D}$ represent the time series from daily reports of new infections, hospitalizations, and deaths, respectively. In addition, $\tau_M$ is the mean time of onset to hospitalization and $\tau_D$ is the mean time of hospitalization to death. We set $\tau_M = 1$, approximating the median value found in \cite{Incubation1}, and the parameter $\tau_D$ is set to $\tau_D = 1$, obtained empirically in the numerical tests.  
Notice that we are not considering the curves of daily reports of ICU admissions, since this is not available in the NYC dataset.

\subsubsection*{Including Gender into the Model}
Covid-19 affects male and female individuals differently. Depending on the age range, the case fatality ratio is much larger for male individuals \cite{gender1,gender2}. To account for gender variance into the model, the transmission parameters ($\beta_M$, $\beta_H$, and $\beta_I$) are generalized. The transmission matrix for the mild class assumes the form:
\begin{equation}
\beta_M = 
\left[
\begin{array}{cc}
    \beta^F_M & \frac{1}{2}(\beta_M^F + \beta_M^M) \\
     \frac{1}{2}(\beta_M^F + \beta_M^M) & \beta^M_M
\end{array}
\right],
\end{equation}
where $\beta_M^F$ and $\beta_M^M$ are the transmission matrices for age ranges, defined above, for female and male individuals, respectively. The transmission matrices $\beta_H$ and $\beta_I$ have the form
$\beta_H = a\beta_M$ and $\beta_I = b\beta_M$.

Notice that the transmission between genders is accounted by the mean value of the transmission inside genders. Intuitively, it means this means that a female individual that keeps social distance with other females will also keep such containment measures with male individuals. The same happens with male individuals.
\subsubsection*{Including Geographical Information}
Monitoring and forecasting the disease spread and the effectiveness of containment measures in large regions, like metropolitan areas, states, and countries, are difficult tasks. Heterogeneous distribution of population and differences in the implementation of social restrictions may lead to quite different disease dynamics from place to place. Moreover, people moving between regions can also cause new infection waves. Thus, to account for these aspects, an epidemiological model must include the population's geographical distribution. A number of approaches have been proposed and a review on this subject can be found in \cite{keeling2008}. Among them, SEIR-type models have been recurrently used to describe the dynamics of human infectious diseases with geographical information. For example, \cite{gatto2020,calvetti2020} describe the Covid-19 dynamics in the United States counties and in Italy, respectively.

In order to include the geographical distribution of the population into the model described by Eqs.~(\ref{eq:modela})-(\ref{eq:model}), we enlarge the transmission matrices. By indexing each site under consideration by $l=1,\ldots,m$, let $\beta_M^l$, $\beta_H^l$, and $\beta_I^l$ be the corresponding transmission matrices. Then, the transmission matrix for mildly infective individuals in the model becomes:

$$
\beta_M = \left[
\begin{array}{ccccc}
 \beta_M^1 & c_1 \textbf{1} & c_1 \textbf{1} & \cdots & c_{1} \textbf{1}\\
 c_1 \textbf{1} & \beta_M^2 & c_2 \textbf{1} & \cdots & c_{2} \textbf{1}\\
 \vdots & \ddots & \ddots & \cdots & \vdots\\
 \vdots & \ddots & \ddots & \ddots & \vdots\\
 c_{1} \textbf{1} & \cdots & c_{m-2} \textbf{1} & c_{m-1} \textbf{1} & \beta_M^m
\end{array}
\right],
$$
where $\textbf{1}$ is the $m\times n$-matrix with all entries equal to one and $c_l = \min_{i,j}  [\text{B}^l]_{i,j}$, 
where $\text{B}^l$ is the matrix defining the transmission matrix $\beta^l_M$, for the $l$-th location. The transmission matrices for the other infective classes are similar. 

This choice for the matrix that represents the mixture of infective populations from different locations helps to simplify the model, reducing considerably the number of unknowns, thus making calibration easier. In addition, the model structure is data driven in the sense that it depends on the current information, thus reflecting more precisely population behavior under different contention measures.

When dealing with large places, like states and countries, it is worth to add to the model the distance between places by using exponential, Gaussian, or power law functions \cite{keeling2008}. 
Due to the interconnectedness of NYC and its boroughs, 
we prefer to estimate the transmission-matrix components from the reported data.

\subsection{Estimation Procedure}
For simplicity, we start by presenting the estimation procedure for the model without gender or geographical dependence. Moreover, the data on new infections, new hospitalizations, and new deaths released by the NYC authorities do not include gender or age-ranges. Thus, we shall use this simpler version of the model, where $\beta^M = \beta(t)\text{B}$. In addition, to simplify the estimation, the constants $a$ and $b$, related to the transmission matrices of hospitalized and in ICU individuals, are empirically set as $a=0.1$ and $b=0.01$, respectively.

In order to estimate the model parameters from publicly available curves of daily new infections, we  build the so-called posterior distribution relating parameters to data. 

We assume the number of daily new infective cases, denoted by $\mathcal{I}$, is Poisson-distributed with parameter $\sigma\sum_{i=1}^n E^i(t)$. Thus, denoting the vector of model parameters by $\theta$, the logarithm of the likelihood function is 

$$
L(\theta) \propto \sum_{j=1}^N\left[\mathcal{I}(t_j). \log\left(\sigma\sum_{i=1}^nE^i(t_j)\right)
\sigma\sum_{i=1}^nE^i(t_j) - \log(\mathcal{I}(t_j)!)\right],
$$
where $N$ is the number of samples in the data and the $\log(\mathcal{I}!)$ is approximated by the Stirling's formula

$$
\log(\mathcal{I}!)\approx \frac{1}{2}\log(2\pi\mathcal{I}) + \mathcal{I}\log(\mathcal{I}) - \mathcal{I}.
$$
We also assume that the vector of parameters $\theta$ is Gaussian{-distributed} with the mean given by some vector of suitably chosen {\it a priori} parameters, denoted by $\theta_0$, 
and identity covariance matrix.
Thus, the negative of the logarithm of the posterior distribution $lP(\theta|\mathcal{I},\theta_0)$ satisfies 
\begin{equation}\label{eq:def_lP}
lP(\theta|\mathcal{I},\theta_0) \propto L(\theta|\mathcal{I}) + \frac{\alpha}{2}\|\theta-\theta_0\|^2.
\end{equation}
The constant $\alpha$ is the so-called regularization parameter in the context of Tikhonov-type regularization methods \cite{ern}. 
The estimated parameters are obtained by minimizing $lP(\theta|\mathcal{I},\theta_0)$.

We estimate the initial proportion of mild infective individual on each age range as follows:
$$
\textbf{I}_M(0) = [I^1_M(0),\ldots,I^n_M(0)]^T \approx I_{M,0}[p_1,\ldots,p_n]^T,
$$
where $I_{M,0}$ is a scalar and $p_i$ is the population fraction of infective individuals on the $i$-th age range. 
The latter is estimated from census data.
Thus, the vector of parameters to be estimated assumes the form:
\begin{equation}
 \theta = [I_{M,0},\beta(t),b_1,\ldots,b_{n-1}]^T,
\end{equation}
where the values of $b_j$, $j~=~1,\ldots,n-1$, correspond to the entries of the matrix $\text{B}$ in Eq.~(\ref{eq:matrixB}). The time-dependent transmission coefficient $\beta(t)$ also appears in the definition of $\beta_M$.

The estimation of $\theta$ proceeds as follows:
\begin{enumerate}
 \item Assume that $\beta(t)$ is constant, and estimate $\theta$ from the set of daily reports of new infections;
 \item Estimate $\beta(t)$ for each $t_j$ in the dataset
 by minimizing the following functional:
\begin{multline}
F\left(\beta(t_{j+1})|\beta(t_j),\theta,\mathcal{I}(t_{j+1})\right) = \mathcal{I}(t_{j+1}).\log\left(\sigma\sum_{i=1}^nE^i(t_{j+1})\right)
- \sigma\sum_{i=1}^nE^i(t_{j+1})\\ - \log\left(\mathcal{I}(t_{j+1})!\right) + \alpha\left(\beta(t_{j+1})-\beta(t_{j})\right)^2,\quad\mbox{with $j=1,\ldots,N-1$.}
\label{eq:post1}
\end{multline}
\end{enumerate}
The estimation of the model with geographical information goes as follows: Consider the $m$-dimensional time series of daily new reported infections from $m$ different places, where $\mathcal{I}^l$, $l=1,\ldots,m$, denotes the set of reports for the $l$-th place. For each $l$, let $\theta^l$ and $\beta^l(t)$ denote the vector of parameters and the time-dependent transmission coefficient, respectively, of the $l$-th place. We estimate the set of parameters $\Theta = [\theta^1,\ldots,\theta^m]$ and the coefficients $\vec{\beta}(t) = [\beta^1(t),\ldots,\beta^m(t)]^T$, by minimizing the log-posterior densities below:

\begin{equation}
    \sum_{l=1}^m lP(\theta^l|\mathcal{I}^l,\theta^l_0), 
    \quad\mbox{ and }\quad
    \sum_{l=1}^m F(\beta^l(t_{j+1})|\beta^l(t_j),\theta^l,\mathcal{I}^l(t_{j+1})),~j=1,\ldots,N-1,
    \label{eq:post2}
\end{equation}
where $lP(\theta^l|\mathcal{I}^l,\theta^l_0)$ and $F(\beta^l(t_{j+1})|\beta^l(t_j),\theta^l,\mathcal{I}^l(t_{j+1}))$ are given by \eqref{eq:def_lP} and \eqref{eq:post1}, respectively. The computational cost of estimating this model varies with the number $m$ of sites considered. Based on the available computational resources, for large m, it may be worth to simplify the model by reducing the number of age ranges or merging sites into a larger one,  thus reducing the model's dimension.

We implemented the model's solution and the estimation procedures in MATLAB\textsuperscript \textregistered. The code is available upon request. The optimization of the posterior density was performed by the general-purpose gradient-based algorithm LSQNONLIN from MATLAB\textsuperscript \textregistered's Optimization Toolbox.

\section{Results: Estimation, Backtesting, and Forecasting}
We start by evaluating the accuracy of the proposed methodology on fitting real data. Then, we do back-testing of the estimated results with out-of-sample data for periods of 7 and 20 days. Finally,  we perform a forecast analysis under different scenarios. 

Our results are based on New York City reports of new infections, hospitalizations, and deaths. This dataset is updated daily and it contains information about the disease distribution at the five NYC boroughs with age-structured accumulated numbers \cite{NYCgithub}.

As the number of daily Covid-19 tests have been increasing in NYC and more effective treatments have been tested \cite{whoCortico}, the rates of new hospitalizations and new deaths have been decreasing, in comparison to the rates of new infections. In order to account for all these features, the rates used in the model were updated.

\subsection{Estimation Results}

Our initial example set does not consider geographical dependence yet. 
Such information shall be incorporated along Subsection~\ref{sec:B}.
The disease dynamics is estimated from the number of daily new infections in the entire NYC area. The data series goes from  29-Feb-2020 to 21-Aug-2020 and comes from publicly available data in \cite{NYCgithub}. This source provides Covid-19 case reports and statistics for NYC and each of its five boroughs. The populations of NYC and of its boroughs is distributed in the 5 age ranges present in the datasets. The population distribution on age ranges and boroughs is based on the census data publicly available at \cite{BaruchColege}. The curves of daily reports of new infections were smoothed-out by a moving average
of seven consecutive days. The rates per 100,000 inhabitants were used to define the various model parameters. They include the  hospitalization, recovery, and death rates, as well as the vector $\textbf{a}$ in the definition of the transmission matrices.

\begin{figure*}[!htb]
  \centering
  \includegraphics[width=0.49\textwidth]{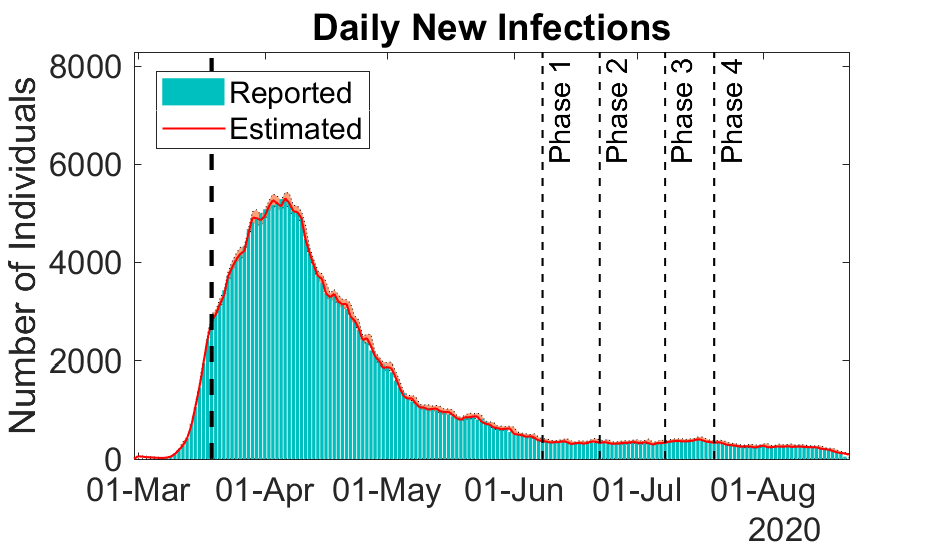}\hfill
      \includegraphics[width=0.49\textwidth]{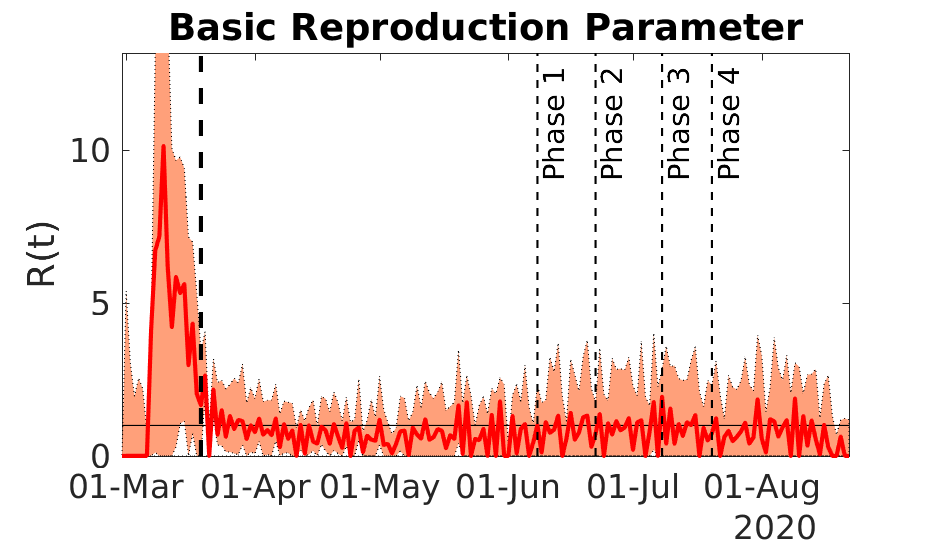}\hfill
     \caption{Model predictions and reported daily new infections (left) and Time-dependent basic reproduction rate (right).
     Solid lines represent best fit predictions and bars are the 7-day moving average of reported cases. Filled envelopes are the 90\% CIs. The vertical dashed lines mark different events in disease dynamics. The first one divides the dataset into uncontained and contained spread and the remaining ones mark the beginning of reopening phases. NYC data.}
  \label{fig:NYC1}
\end{figure*}

After preliminary calibration tests, two disease evolution regimes were clearly identified. In the first one, the number of infective individuals increased exponentially and, in the second one, the spread was considerably reduced due to the contention measures imposed by the state of emergency declared on 12-Mar-2020. The effect of such intervention was clearly observed in the evolution of the time-dependent transmission coefficient $\beta(t)$ on 19-Mar-2020 (Fig.~\ref{fig:NYC1}). 

In order to capture possible regime changes, like the different age-range mixing, we divided the time-series into two parts, namely data before and after 19-Mar-2020. For these two time series subsets, we estimate the vector of parameters $\theta$ and the time-dependent $\beta$. Note that, for the second dataset, after 19-Mar-2020, we do not estimate the initial infective population. The estimated parameters and the corresponding 90\% confidence intervals (CIs) can be found in Tab.~\ref{tab:NYC1}. Such intervals were obtained by excluding the 5\% largest and smallest values generated by 200 bootstrap samples \cite{chowell2017}. Figure~\ref{fig:NYC1} presents the comparison between reported and model predicted curves for daily new infections. 
The time evolution of the basic reproduction rate can be found in Fig.~\ref{fig:NYC1}. Table~\ref{tab:NYCreports} depicts the predicted and reported number of infections, hospitalizations, and deaths for 21-Aug-2020 with 90\% CIs.

\begin{table}[!htb]
\centering
\begin{tabular}{c|c|c}
\cline{2-3}
          & Subset 1                     & Subset 2   \\
\hline
$I_{M,0}$ & 2.0 (2.0--2.0) &   -\\
$\beta$   & 8.29 (8.15--8.35)  &  2.81 (2.71--3.20)\\
$b_1$     & 0.88 (0.87--0.89)  &  0.51 (0.50--0.81)\\
$b_2$     & 0.81 (0.79--0.81)  &  0.24 (0.01--0.31)\\
$b_3$     & 0.77 (0.75--0.78)  &  0.19 (0--0.53)\\
$b_4$     & 0.52 (0.32--0.58)  &  0.17 (0.11--0.85)\\
\hline
\end{tabular}
\caption{Best fit and 90\% CIs of the model parameters obtained from the time-series subsets of daily reported new infections in NYC.}
\label{tab:NYC1}
\end{table}

In Tab.~\ref{tab:NYC1}, the estimated values of $\beta$ were far from each other in both periods, which indicates the aforementioned change in the transmission regime. Concerning the vector $\textbf{b}$, after contention, the estimated values decrease consistently, indicating that the interaction between age ranges was also significantly reduced. All these results show that, after 19-Mar-2020, disease spread was contained. However, as we shall see later on, new infection waves can occur if contention measures are relaxed.

\begin{table}[!htb]
\centering
\begin{tabular}{c|ccc}
\hline
Age & Infections & Hospitalizations & Deaths\\
\hline
0-17 & 7443 (6904--10050) & 610 (566--822) & 11 (10--15)\\
& 7252 & 623 & 12\\
\hline
18-44 & 88526 (82183--119054) & 9175 (8520--12321) & 716 (665--966)\\
& 86413 & 9297 & 728\\
 \hline
45-64 & 83391 (77466--111818) & 18712 (17388--25054) & 4217 (3923--5675)\\
& 81125 & 18958 & 4268\\
 \hline
65-74  & 26131 (24271--35046) & 12856 (11947--17216) & 4889 (4557--6587)\\
&27221 & 12434 & 4707\\
\hline
75+ & 23098 (21456--30947) & 16712 (15534--22356) & 10070 (9408--13569)\\
&25653 & 15568 & 9298\\
 \hline
Total & 228588 (212280--306916) & 58065 (53954--77769) & 19902 (18564--26812)\\
&228144 & 56882 & 19014\\
\hline
\hline
Gender & Infections & Hospitalizations & Deaths\\
\hline
Female & 114649 (106465--153971) & 24215 (22499--32443) & 7212 (6725--9719)\\
&111713 & 24934 & 7638\\
\hline
Male & 113939 (105815--152945) & 33850 (31455--45326) & 12690 (11839--17094)\\
&116239 & 31936 & 11373\\
\hline
\end{tabular}
\caption{Model predictions with 90\% CIs (top rows) and reported numbers (bottom rows) of accumulated infections, hospitalization, and deaths for NYC on 21-Aug-2020, by age range and gender.}
\label{tab:NYCreports}
\end{table}

Figure~\ref{fig:NYC1} shows the adherence of the calibrated model predictions to the 7-day moving average of the reported number of new infections. 
The hospitalization and death rates were evaluated using the ratios of reported data defined previously.  
The proportions of ICU admissions by age were obtained in \cite{covid2020severe} and they were adjusted according to the proportions of deaths available in \cite{NYCgithub}. The model accuracy is also illustrated in Tab.~\ref{tab:NYCreports}, where we can see the model predictions for the accumulated numbers of infections, hospitalizations, and deaths, for each age range and gender, are close to the reported ones.

As Fig.~\ref{fig:NYC1} clearly shows, the time-dependent basic reproduction rate $R(t)$ has two different levels. Before 19-Mar-2020, it presents large values, indicating that the disease was spreading without control. After 19-Mar-2020, the transmission parameter value drops to around one, which indicates control of transmission by contention measures imposed from 12-Mar-2020 onwards. Notice that, the large values for the basic reproduction rate in the first part of the series, i.e., before 19-Mar-2020, might be caused by an accumulation of reports in the beginning of the outbreak. Such accumulation can be related, for example, to difficulties faced by the health authorities to setup an appropriate diagnosis protocol.

The adherence of the calibrated model predictions to reported data, the accuracy in the number of hospitalizations and deaths, as well as the behavior of the calibrated parameters indicating the effect of disease contention measures for NYC data show that our proposed model captures well the Covid-19 dynamics in NYC. Therefore, it is useful to track the spread dynamics, allowing to assess the effects of, for example, travel quarantine, social distancing, and reopening strategies. If infection curves for different age ranges are available, it is possible to use the present model to track aspects like the effects of reopening schools, universities, or yet parks and public gardens, since these spaces are usually frequented by people of well defined age ranges. Thus, it is easier to track the disease spread dynamics more accurately, allowing the decision of whether additional reopening or further restrictions could be taken.

\subsection{Including Geographical Information}\label{sec:B}

Let us consider the epidemiological dynamics of the five boroughs of NYC, namely, Queens, Manhattan, Staten Island, Brooklyn, and Bronx. The data was downloaded from \cite{NYCgithub} on 02-July-2020.

\begin{figure}[!htb]
  \centering
      \includegraphics[width=0.4\textwidth]{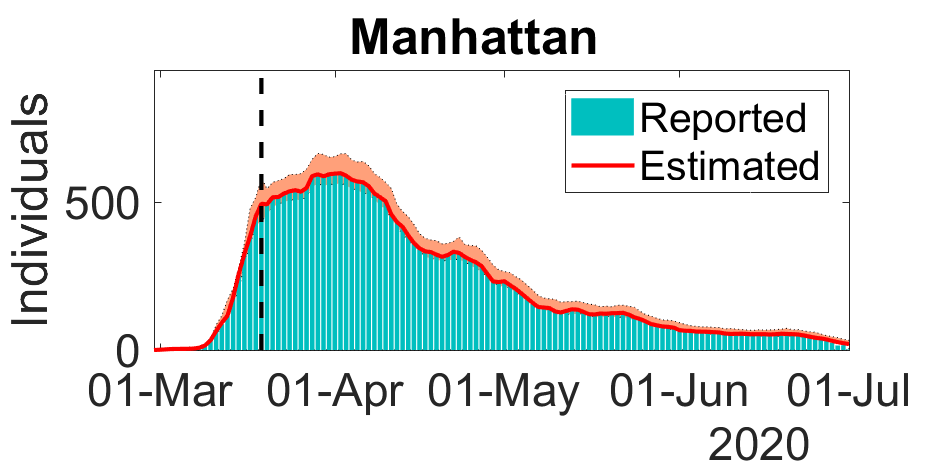}\hfill
      \includegraphics[width=0.4\textwidth]{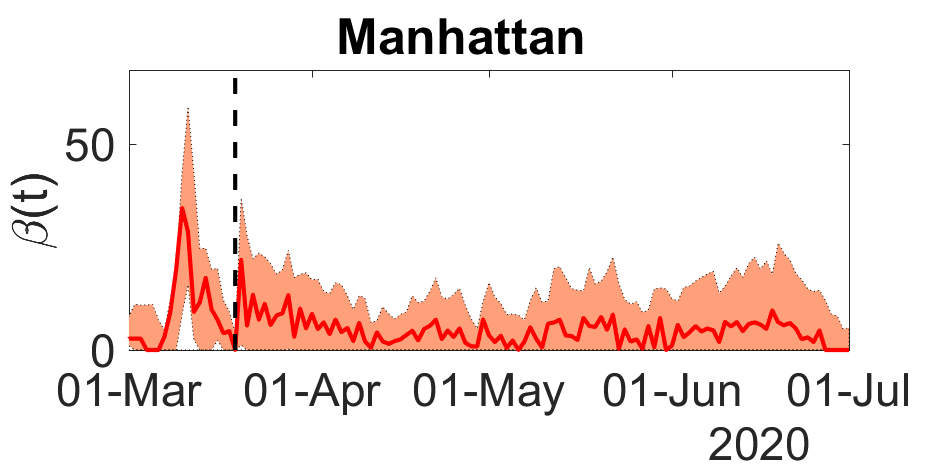}\hfill
      \includegraphics[width=0.4\textwidth]{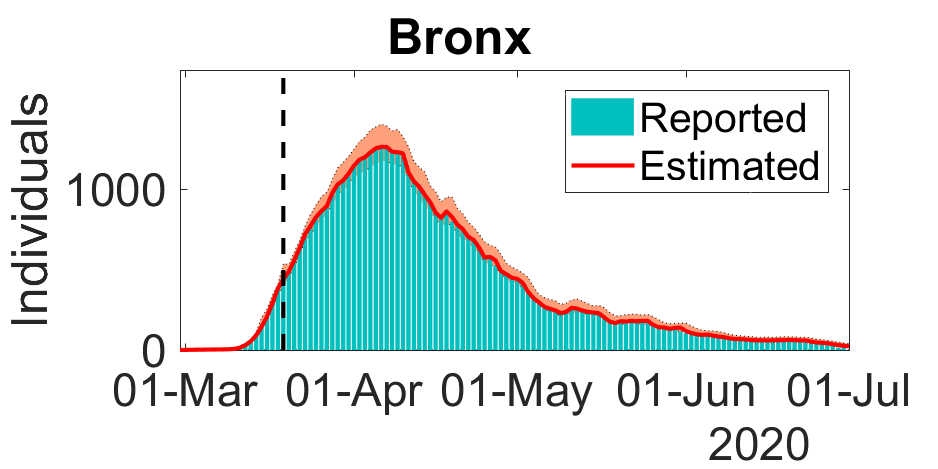}\hfill
      \includegraphics[width=0.4\textwidth]{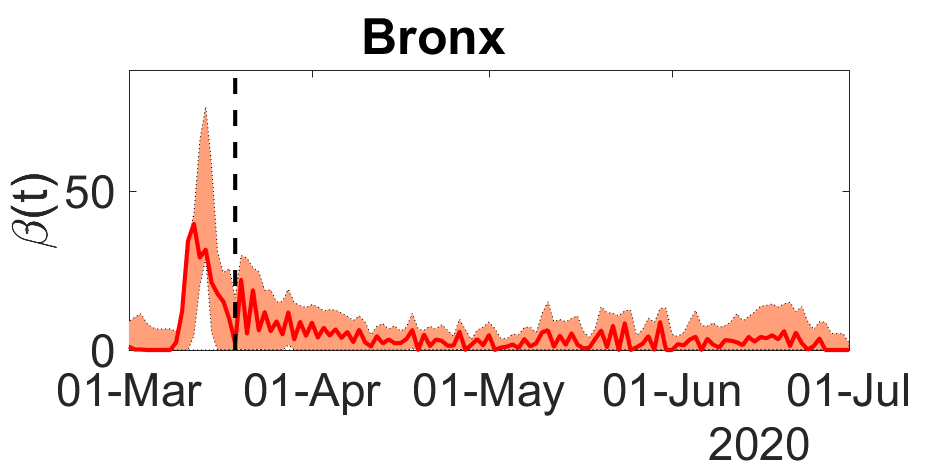}\hfill
      \includegraphics[width=0.4\textwidth]{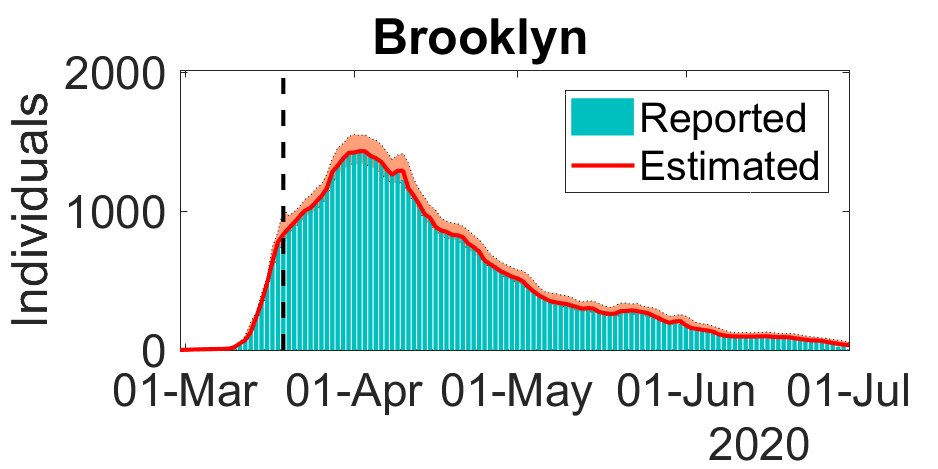}\hfill
      \includegraphics[width=0.4\textwidth]{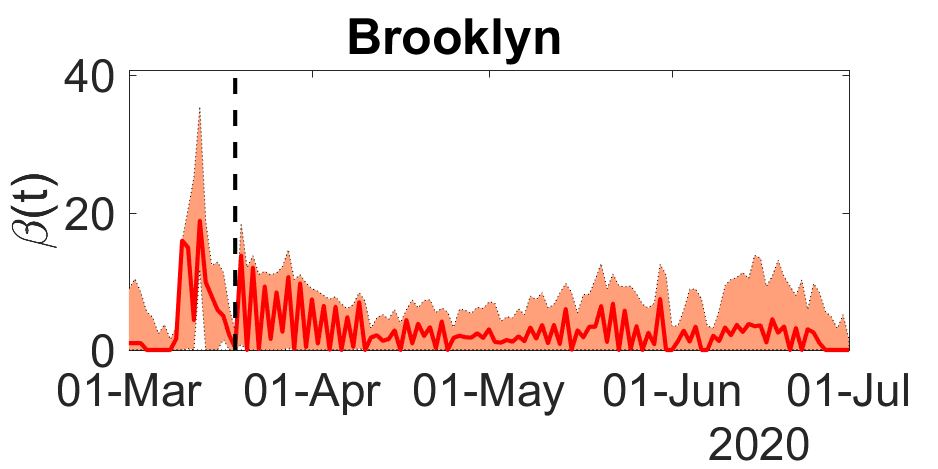}\hfill
      \includegraphics[width=0.4\textwidth]{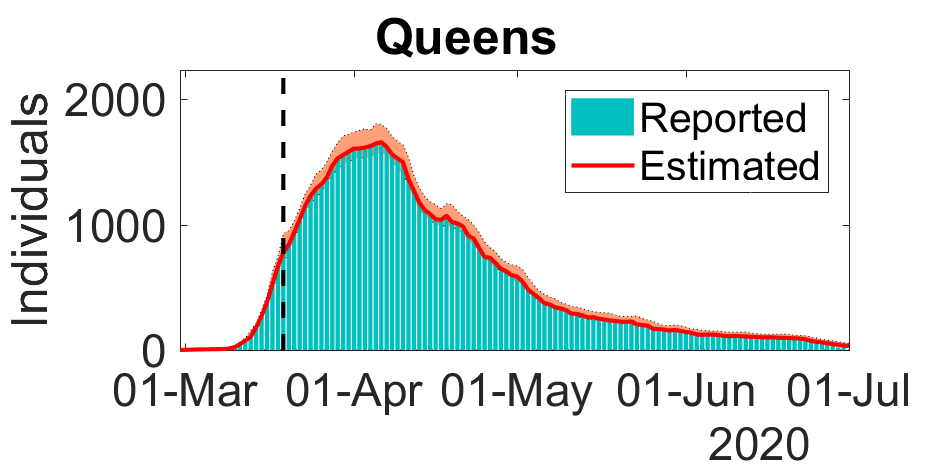}\hfill
      \includegraphics[width=0.4\textwidth]{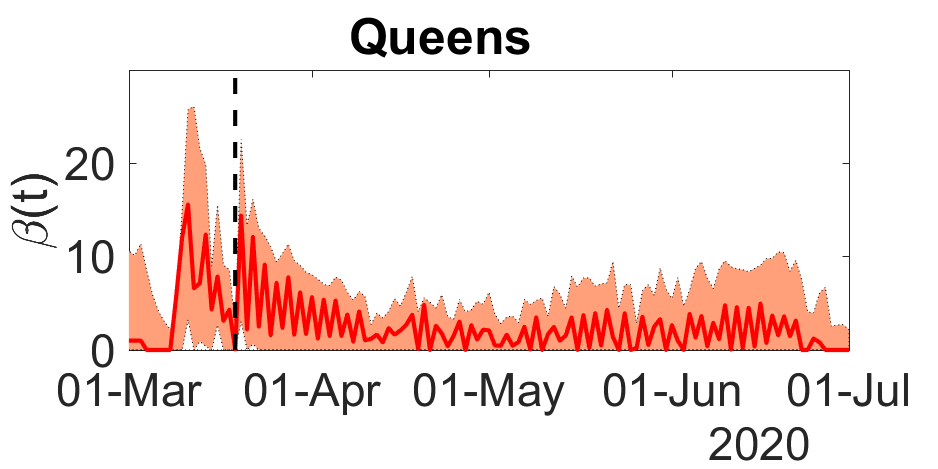}\hfill
      \includegraphics[width=0.4\textwidth]{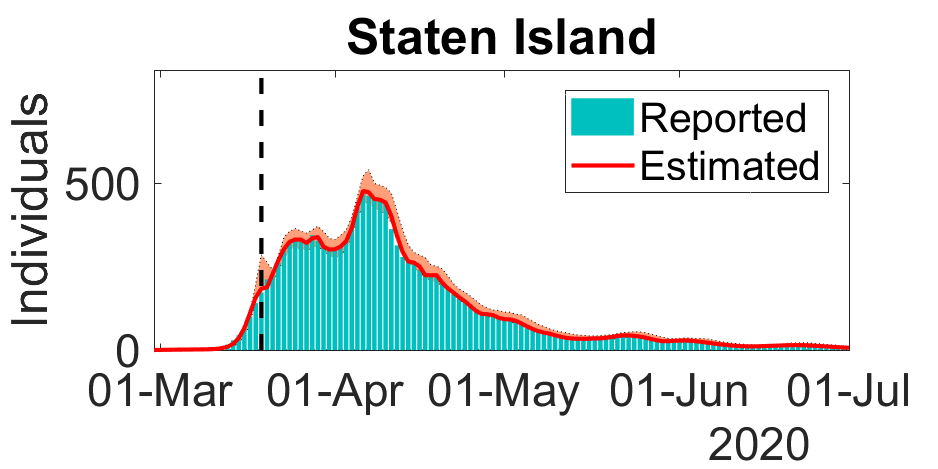}\hfill
      \includegraphics[width=0.4\textwidth]{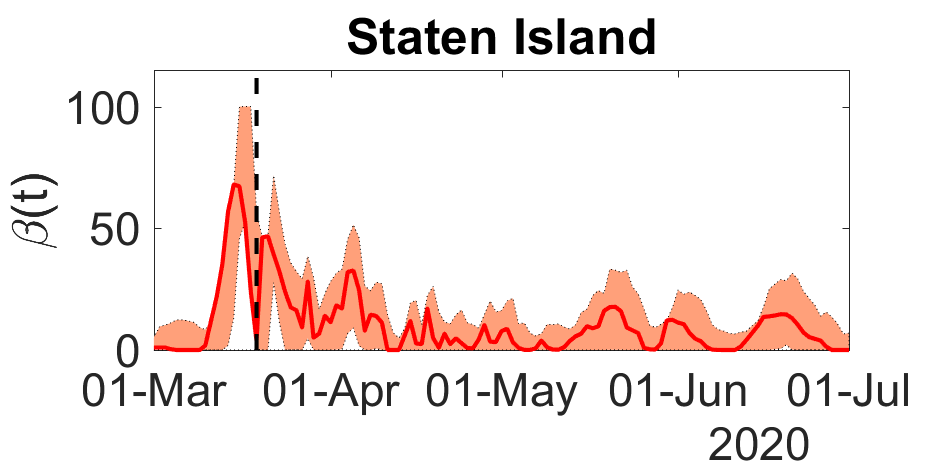}
     \caption{Left Column: Model predictions and reported daily new infections for each NYC borough. Right Column: Time-dependent transmission parameter estimated for the NYC each boroughs. Solid lines represent the model best fit predictions and bars depict the 7-day moving average of the number of reported cases. Dashed lines divides the spread regime into uncontained and contained. The filled envelopes are the 90\% CIs. Time series starts on 29-Feb-2020 to 01-July-2020.}
  \label{fig:boro1}
\end{figure}

The calibrated model predictions of daily new infections and the corresponding time-dependent transmission coefficients can be found in Fig.~\ref{fig:boro1}.

\begin{table}[!htb]
\centering
\begin{tabular}{c|ccc}
\hline
Borough &  Infections & Hospitalizations & Deaths\\
\hline
Manhattan & 26886 (26793--28994) & 8560 (8540--9269) & 2826 (2805--3052)\\
 & 26809 & 8081 & 2448\\
\hline
Bronx & 47870 (46565--51577) & 13618 (13486--15226) & 4882 (4867--5629)\\
 & 47691 & 12233 & 3830\\
\hline
Brooklyn & 59287 (58357--63889) & 14062 (13900--15177) & 4855 (4741--5260)\\
 & 59037 & 15351 & 5495\\
\hline
Queens & 64962 (63939--70092) & 15571 (15318--16786) & 4917 (4829--5270)\\
 & 64797 & 16977 & 5839\\
\hline
Staten Island & 14131 (14131--15421) & 2287 (2287--2523) & 840 (840--927)\\
 & 13961 & 2351 & 872\\
\hline
\end{tabular}
\caption{Model predictions and reported accumulated infections, hospitalizations, and deaths for the boroughs of NYC on 01-July-2020. 
Top rows represent predictions and bottom rows are reported cases. 
Inside the parentheses are the 90\% CIs.}
 \label{tab:boro}
\end{table}

Figure~\ref{fig:boro1}, among other things, shows the model predictions adherence to the curves of reported daily new infections. Such accuracy is also attested by the comparison between reports and predictions of the accumulated number of total infections, hospitalizations, and deaths for 01-July-2020 in Tab.~\ref{tab:boro}. 

The behavior of the time-dependent transmission coefficient for each borough is similar to the basic reproduction rate $R(t)$ in the previous example. This behavior is expected since transmission contention measures were taken since 12-Mar-2020 in the entire NYC. 

This example shows the ability of the present model to detect disease transmission patterns in different locations at the same moment.
The model also accounts for the interaction between individuals from different age ranges, genders, and locations. With such features, it is possible to track the implications of reopening, the necessity of additional contention measures, or yet auxiliating the design of vaccination strategies. Such broad applications could not be achieved via simpler models.

\subsection{Backtesting}
In order to test short-term forecast capabilities of the model, we consider two different periods of the Covid-19 outbreak in NYC.

{\subsubsection*{Uncontained Spread}}
We calibrate the parameters with data from reports on new infections in the period from 29-Feb-2020 to 19-Mar-2020 and we produce a seven-day forecast starting on 20-Mar-2020. This forecasted period is of particularly interest since on 19-Mar-2020 the disease spread pattern changed considerably due to the contention measures undertaken 7 days earlier, as a consequence of the state of emergency declared on 12-Mar-2020. To generate the predictions, we assume that $\beta(t)$ is constant for dates $t$ after 19-Mar-2020 and that it takes the same value as the one estimated on 19-Mar-2020.

\begin{figure*}[!htb]
  \centering
      \includegraphics[width=0.33\textwidth]{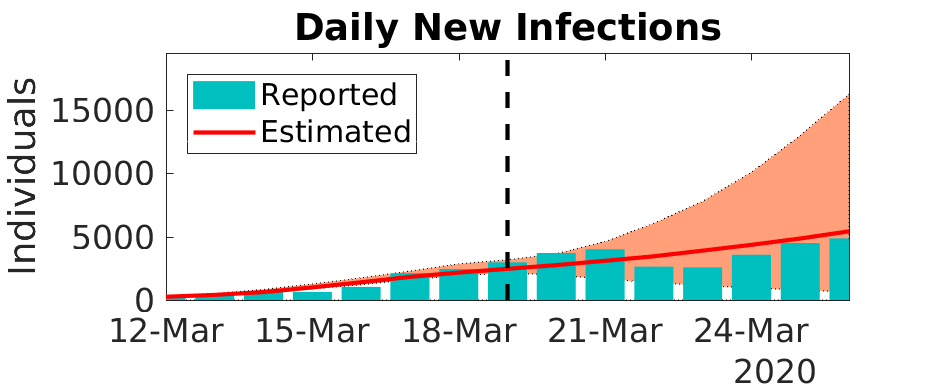}\hfill
      \includegraphics[width=0.33\textwidth]{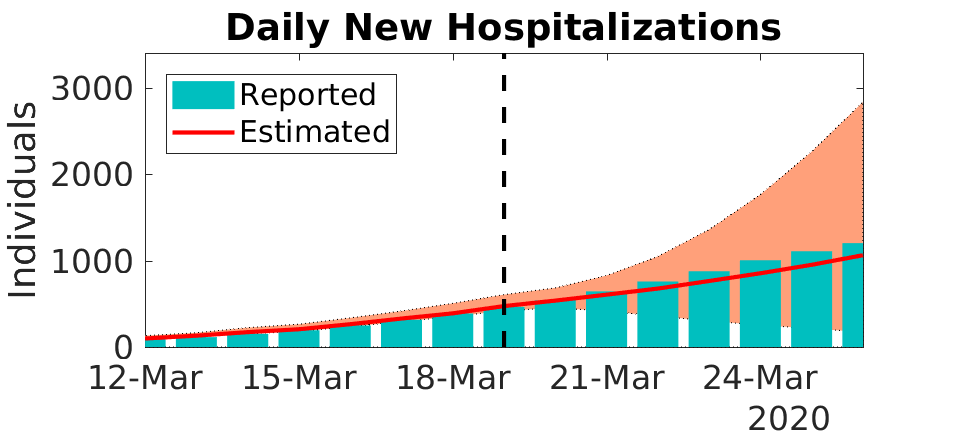}\hfill
      \includegraphics[width=0.33\textwidth]{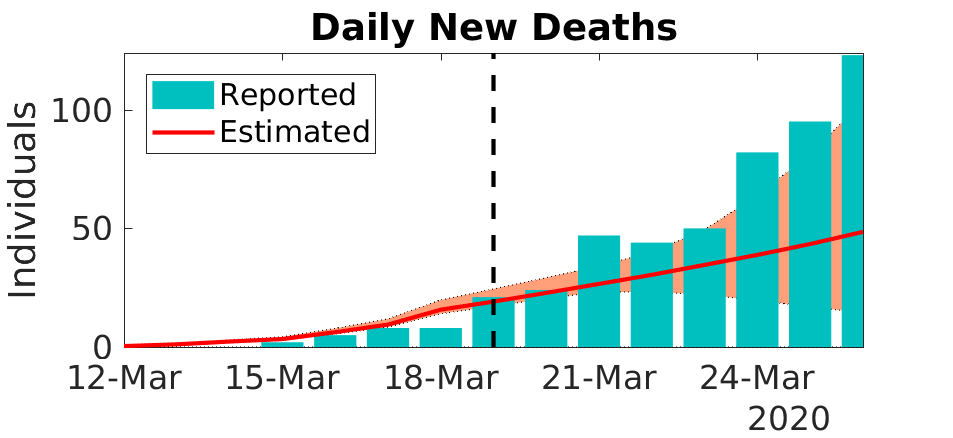}\hfill
            \includegraphics[width=0.33\textwidth]{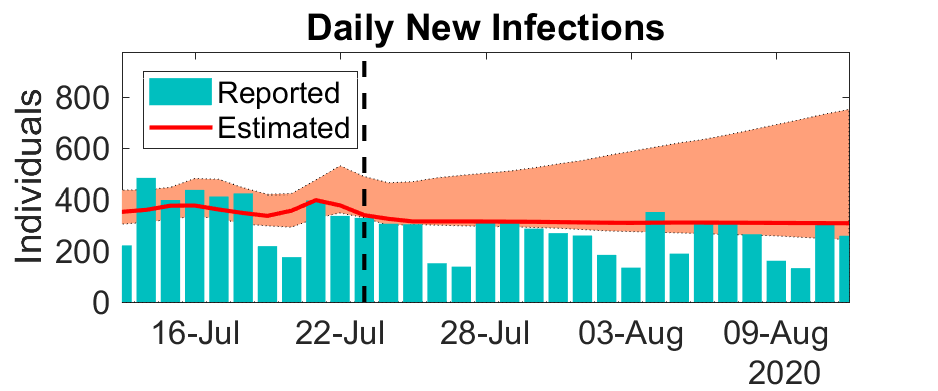}\hfill
      \includegraphics[width=0.33\textwidth]{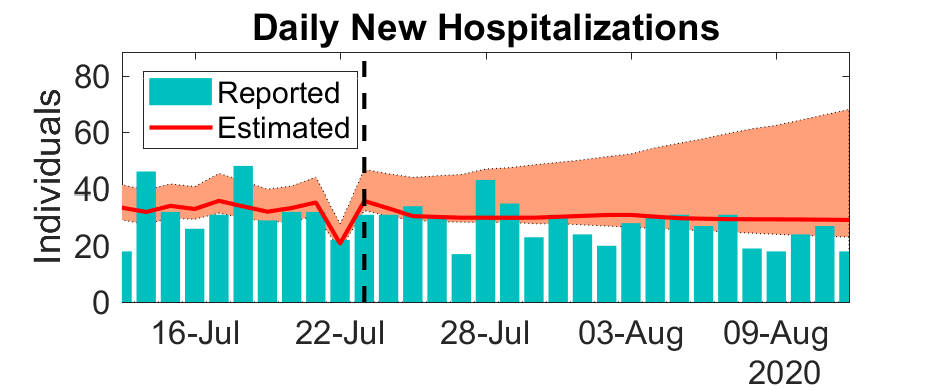}\hfill
      \includegraphics[width=0.33\textwidth]{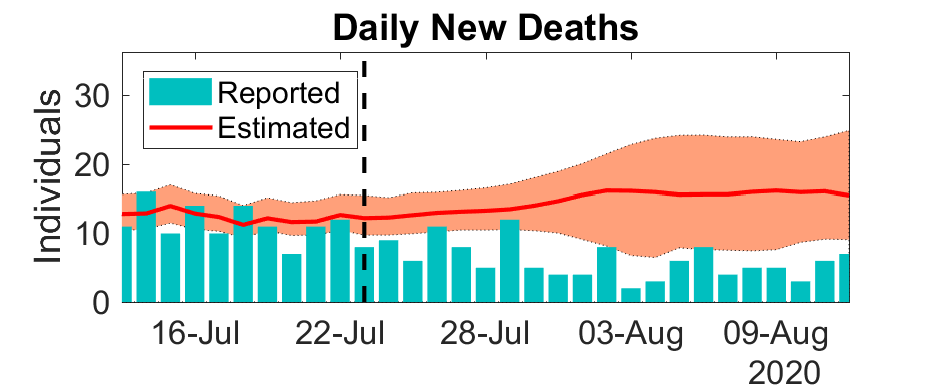}\hfill
     \caption{Predictions for the daily reported new infections (left), new hospitalizations (center), and new deaths (right) for the periods 19 to 24-Mar-2020 (top row) and 24-July-2020 to 12-Aug-2020 (botton row). 
    The solid lines represent the model predictions with best fit and the bars that depict the reported NYC data. On the right side of the dashed lines are the model forecasts. The filled envelopes are the 90\% CIs.}
  \label{fig:forecast1}
\end{figure*}

Figure~\ref{fig:forecast1} presents a comparison between the forecasted curves and the reported data. Although the parameters were estimated with information based on uncontrolled disease spread, the 7-day ahead forecast for daily new infections and new hospitalizations starting on 20-Mar-2020 are satisfactorily accurate. The accumulated number of infections, hospitalizations, and deaths for the forecasted period can be found in Tab.~\ref{tab:forecast2}.

\subsubsection*{Contained Spread}
We now create a forecast for the period 24-July-2020 to 12-Aug-2020. For dates $t$ after 23-July-2020, the time-dependent values of the transmission coefficient are given by the mean of values for the period from 13 to 23-July-2020.

Figure~\ref{fig:forecast1} presents a comparison between the predicted and the reported data for the forecasted period. Table~\ref{tab:forecast2} presents the predicted and the reported values of accumulated infections, hospitalizations, and deaths from 24-July-2020 to 12-Aug-2020.

\begin{table}[!htb]
\centering
\begin{tabular}{c|ccc}
\hline
Period & Infections & Hospitalizations & Deaths\\
\hline
19-Mar--26-Mar & 27838   (8544--61140) &  5456   (2181--10819)  &  245 (143--390)\\
& 25876 & 6141 & 465\\
\hline
24-July--12-Aug & 6262 (5581--11754) & 602 (534--1057)  &  297 (193--393)\\
& 6654 & 736 & 147\\
\hline
\end{tabular}
\caption{Model predictions and reported accumulated infections, hospitalizations, and deaths for the periods
19 to 26-Mar-2020 and 24-July to 12-Aug-2020. Top rows represent predictions while bottom rows are the reported cases.
Between parentheses are the 90\% CIs.}
 \label{tab:forecast2}
\end{table}

As Fig.~\ref{fig:forecast1} and Tab.~\ref{tab:forecast2} show, the model predictions of infections, hospitalizations, and deaths are once again satisfactorily accurate. These results are explained by the model ability to incorporate the disease dynamics through the time-dependent parameters. Notice that, in these examples, the rates of hospitalizations and deaths are evaluated using appropriate ratios of reported data, defined previously, until the last day of estimation. In the forecasted period, we repeat the rates for the days 19-Mar-2020 and 23-July-2020, respectively, for the corresponding  data ranges studied.

\subsection{Reopening Scenarios}
We now apply the calibrated models from the previous sections to a number of plausible scenarios, such as the reopening of the entire NYC region, of schools, and of just one single borough, which we chose to be Staten Island. The predictions for those scenarios will give us an idea of the predictive capability of our model.

\subsubsection*{The Entire NYC}
The aim of this example is to present possible scenarios for the Covid-19 epidemic for long periods without an effective vaccine or appropriate treatment. We consider two different situations. In the first one, the transmission parameters stay at the level of strict containment, what was observed in the period from 04 to 14-June-2020. Thus, for any date $t$ after 14-June-2020, the transmission coefficient $\beta(t)$ is set as the mean for the estimated values of $\beta(t)$ on the period from 04 to 14-June-2020, i.e., $\beta(t) = 1.77$ with 90\% CI 0.27--4.69. In the second case, we simulate a controlled reopening, by allowing the coefficient $\beta(t)$ to reach the double of the values obtained in the previous case, i.e.,  $\beta(t) = 3.54$ (0.53--9.38). However, if we impose that whenever the number of daily new infections reaches 1000 cases, containment measures are undertaken, forcing $\beta(t)$ to return to lower levels until it reaches the value 1.77 (0.27--4.69). On the other hand, after undertaking containment measures, if the number of daily new infections is below 200 cases, we permit reopening, and $\beta(t)$ may grow again until it reaches the value 3.54 (0.53--9.38).

\begin{figure*}[!htb]
  \centering
      \includegraphics[width=0.33\textwidth]{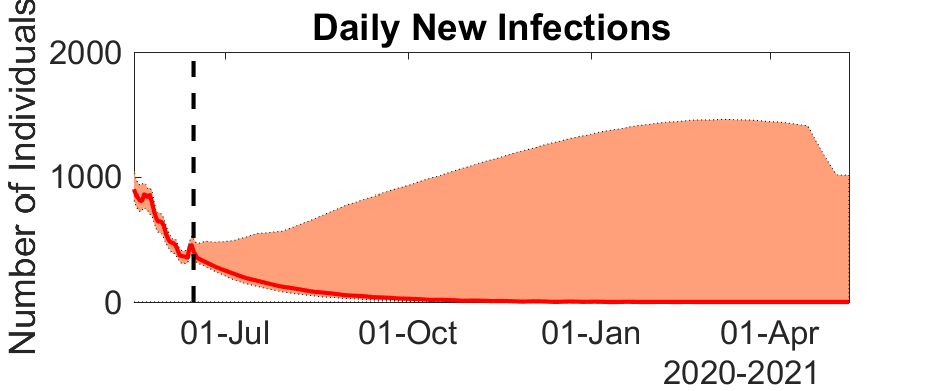}\hfill
      \includegraphics[width=0.33\textwidth]{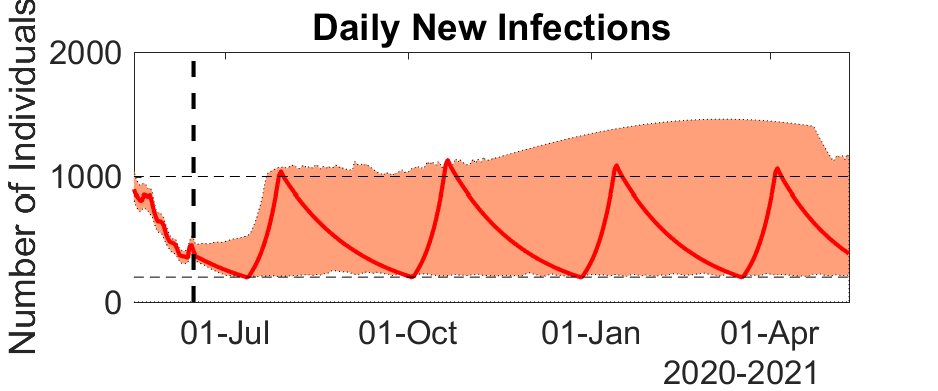}\hfill
      \includegraphics[width=0.33\textwidth]{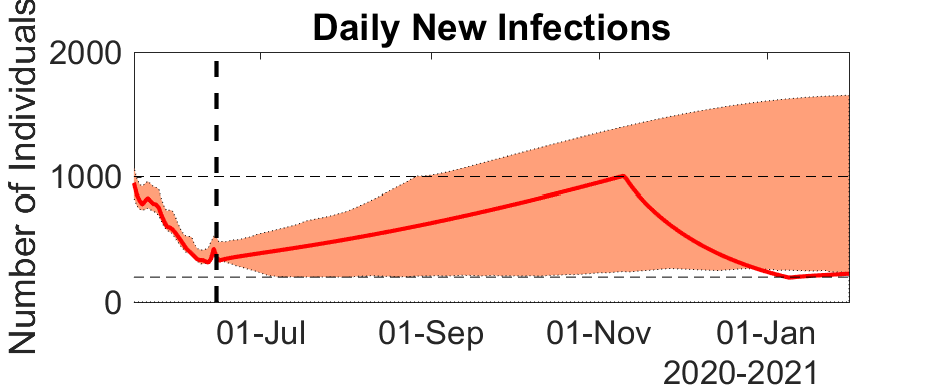}\hfill  
     \caption{Model predictions of daily new infections for reopening scenarios. Left: containment measures are kept during the whole period. Center: Reopening is allowed until the daily number of new infections reaches 1000 or if it is below 200 (horizontal dashed lines). Right: Reopening schools. The filled envelopes represent the 90\% CIs.}
  \label{fig:forecast3}
\end{figure*}

Figure~\ref{fig:forecast3} (left and center panels) present
the curves of daily new infections for the aforementioned  scenarios. Before 14-June-2020, we have the estimated curves. 
The forecasting goes from 15-June-2020 to 11-May-2021. According to this example, the reopening without an effective vaccine or the achievement of herd immunity may bring new infection waves, even when the number of daily new cases is relatively low.

\subsubsection*{Reopening Schools}

To simulate a controlled reopening of schools from 15-June-2020, we use the set of parameters estimated after 19-Mar-2020. In order to artificially increase the interaction between school-age individuals, the entries of the transmission matrix $\beta_M$ associated to the mildly infective individuals in the age range 0 to 17 years old are multiplied by $2.5$. In addition, the time-dependent transmission coefficient is set to $1.25$ times the mean of the estimated values for the period 06 to 15-June-2020. For this specific age range, the transmission parameter values are similar to the ones obtained for the period before 19-Mar-2020. This 
is expected, since controlling the mixing in youth population at schools is difficult. Indeed, recent news indicates that Covid-19 transmission rate amongst people under 19 years old is similar to transmission rate in other age ranges \cite{NYT,CDC2020}.

Figure~\ref{fig:forecast3} (right panel) presents model predictions of this ``safe'' reopening. Notice that, whenever the number of daily new infections reaches 1000 cases, contention measures are imposed again. Reopening occurs if the number of daily new cases would fall below 200.

Therefore, even under an idealized situation, reopening schools may cause new infection waves among the entire population. Thus, monitoring transmission dynamics is of fundamental importance to set the right time for relaxing or tightening contention measures. In Israel, the recent reopening of schools caused a secondary wave of new infections that forced the adoption of new contention measures \cite{NYT}.

\subsubsection*{Reopening Staten Island}

Two different scenarios are considered in this example: reopening Staten Island with and without restrictive measures. In the first one, contention measures are slightly relaxed, without allowing people from different age ranges and boroughs to interact. Quantitatively, we keep the same parameter values estimated in the period 19-Mar-2020 onwards. The only change is in the transmission coefficient for Staten Island, which is set as the double of the mean of the corresponding estimated values of $\beta(t)$ for the period 22-June-2020 to 01-July-2020, for dates $t$ after 01-July-2020. During the forecasted period (02-Jul-2020 to 29-Oct-2020), the transmission coefficients for the other boroughs are kept equal to the mean of the estimated values for the period 22-June-2020 to 01-July-2020.

In the second scenario, people from different boroughs and age ranges are allowed to interact, keeping some social distance and simple containment measures. Only in Staten Island, simple containment measures are undertaken. In other words, the time-independent transmission parameters assume the same values estimated in the period 29-Feb-2020 to 19-Mar-2020. In addition, we allow the time-dependent transmission coefficient for Staten Island to reach the double of the mean of the estimated values for the period 22-June-2020 to 01-July-2020, for dates $t$ after 01-July-2020. Again, after 01-July-2020, the transmission coefficients for the other boroughs are kept equal to the mean of the estimated values for the period 22-June-2020 to 01-July-2020. Whenever the number of daily new infections reaches 1000 in NYC, contention measures are undertaken again. So, the time-independent transmission parameters are brought back to the same values of the period after 19-Mar-2020 and the transmission coefficient for Staten Island is reset to the mean of the corresponding estimated values for the period 22-June-2020 to 01-July-2020. Reopening reoccurs whenever the number of daily new infections in NYC is below 100 reports.

\begin{figure}[!htb]
  \centering
      \includegraphics[width=0.33\textwidth]{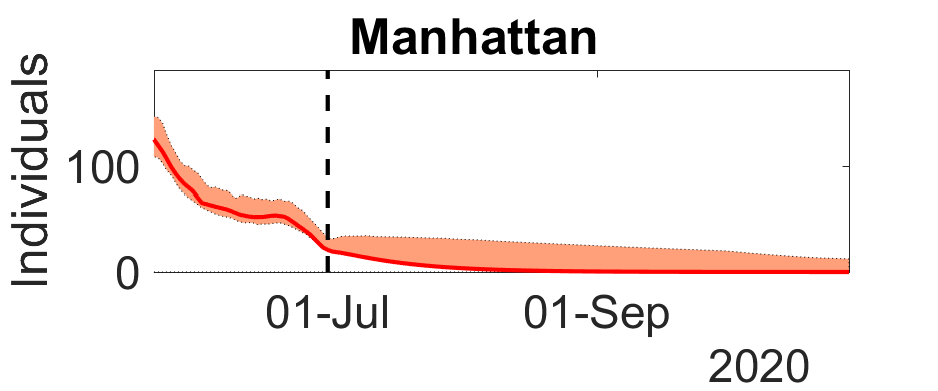}\hfill
      \includegraphics[width=0.33\textwidth]{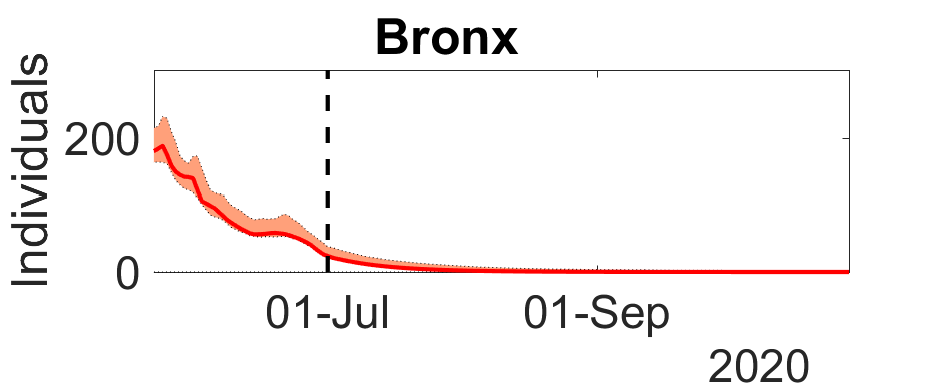}\hfill
      \includegraphics[width=0.33\textwidth]{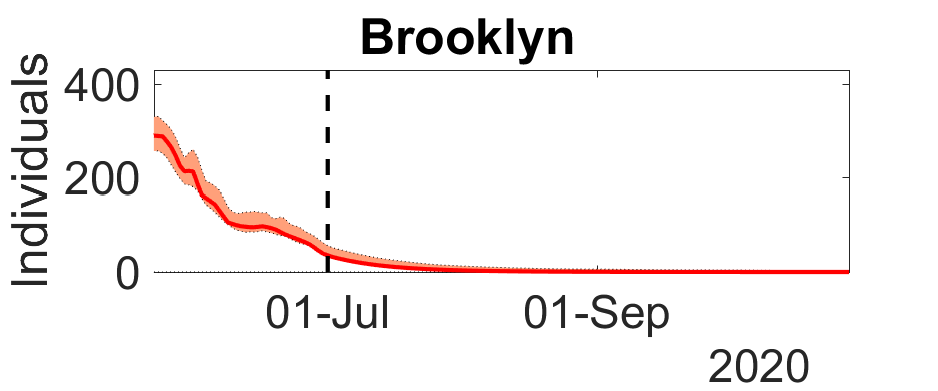}\hfill
      \includegraphics[width=0.33\textwidth]{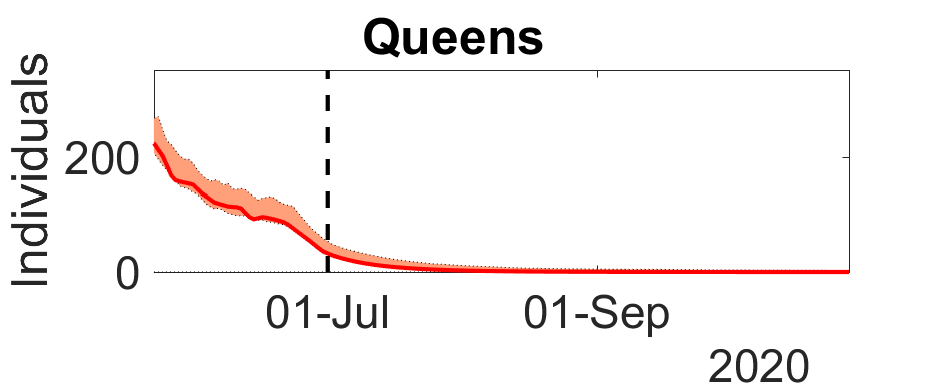}\hfill
      \includegraphics[width=0.33\textwidth]{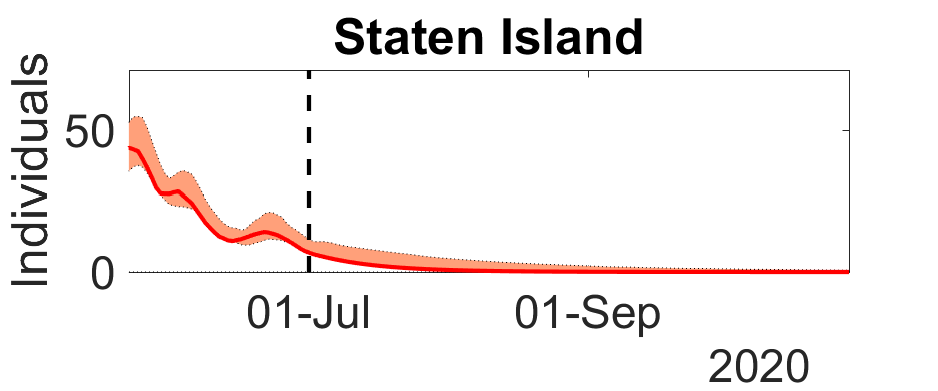}\hfill
      \includegraphics[width=0.33\textwidth]{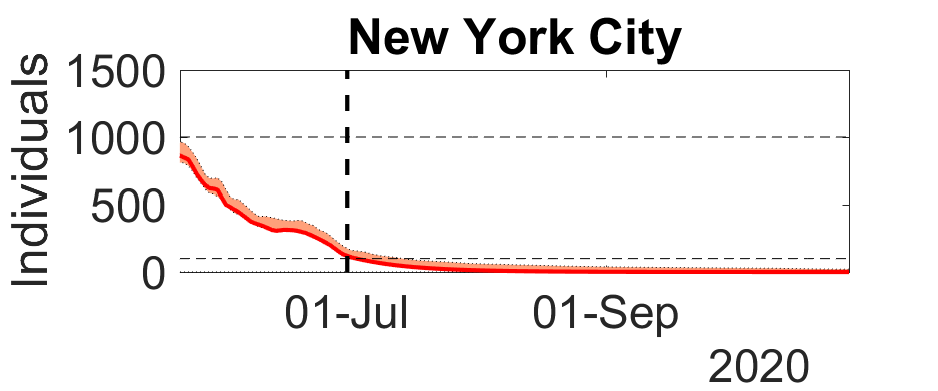}\hfill
     \caption{Model predictions of daily new infections for the period
     02-July-2020 to 29-Oct-2020 in NYC, when lockdown is slightly lifted, but keeping strict containment measures. The filled envelopes represent the 90\% CIs.}
  \label{fig:liftA}
\end{figure}
\begin{figure}[!htb]
  \centering
      \includegraphics[width=0.33\textwidth]{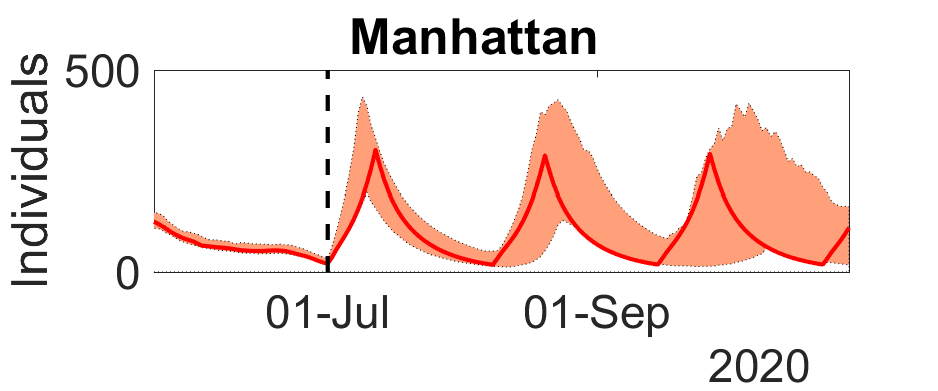}\hfill
      \includegraphics[width=0.33\textwidth]{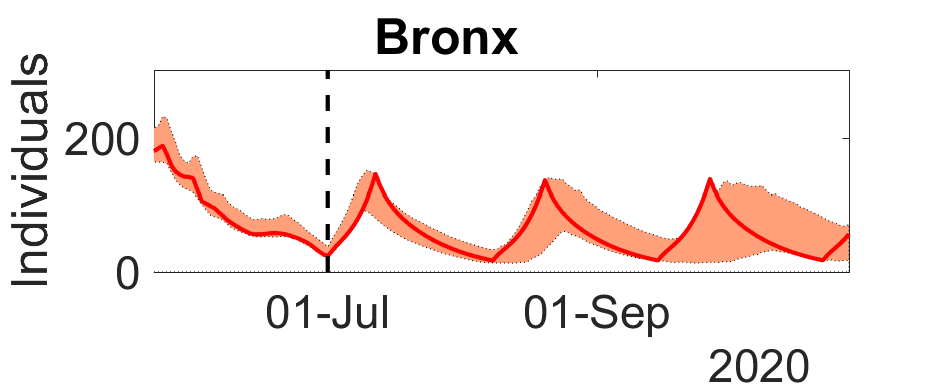}\hfill
      \includegraphics[width=0.33\textwidth]{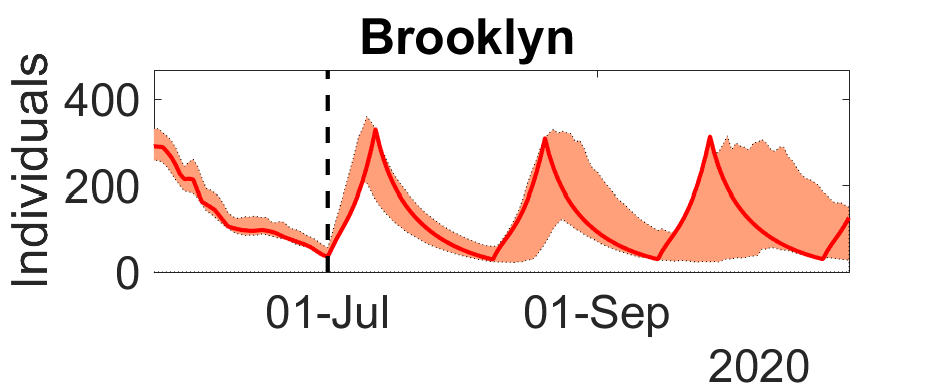}\hfill
      \includegraphics[width=0.33\textwidth]{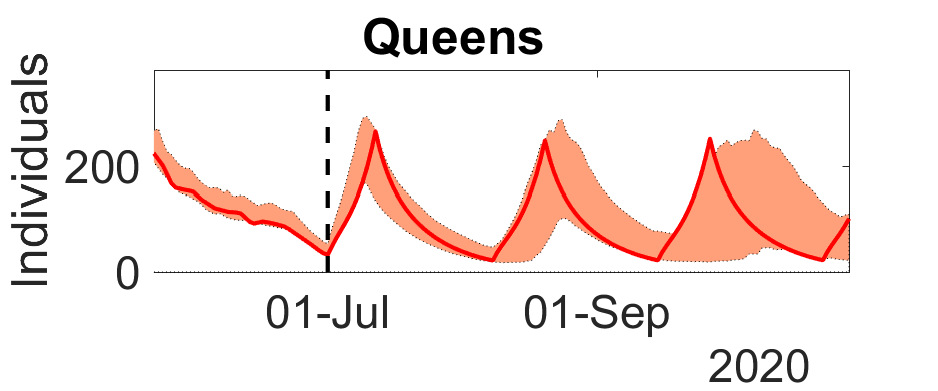}\hfill
      \includegraphics[width=0.33\textwidth]{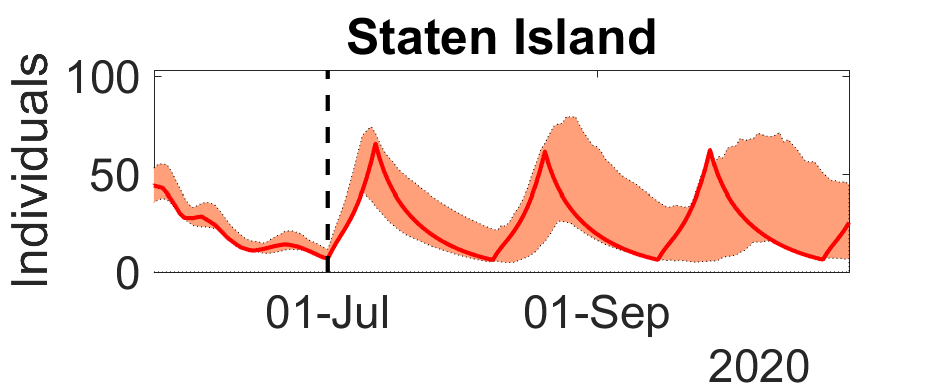}\hfill
      \includegraphics[width=0.33\textwidth]{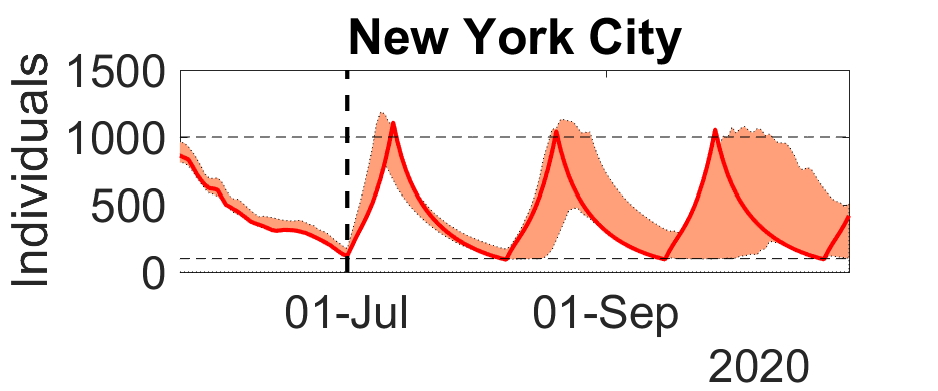}\hfill
     \caption{Model predictions of daily new infections for the period 02-July-2020 to 29-Oct-2020 in NYC, when lockdown is lifted, with light containment measures. The filled envelopes represent the 90\% CIs.}
  \label{fig:liftB}
\end{figure}

Figures~\ref{fig:liftA}-\ref{fig:liftB} present the curves of daily new infections for  the first and second scenarios, respectively. They show the evolution for each borough and for the entire NYC. 
In the first case, doubling the mean of the transmission coefficient alone is not sufficient to cause secondary waves of infection. In other words, if restriction of movement between boroughs are kept and the interaction between age ranges stay contained, through strict social distance measures, infection will not return and the disease outbreak will die out. On the other hand, reopening a borough and allowing people of different ages and from different boroughs to interact, although keeping some light containment and social distance measures, can cause new waves of infection for large periods.

\section{Discussion}
We use a 7-day moving average to perform smoothing of data. After calibration, the model predictions were adherent to the data of daily new infections and they predicted well the number of daily new hospitalizations and deaths. The adjustment of the rates of hospitalization and death using appropriate ratios of reported data contributed to improving model predictions. While backtesting, forecasts for few days ahead, under different contexts, turned out to be accurate. 
The model identified well the effects of lockdown undertaken in NYC after 12-Mar-2020. A considerable change in the values of the transmission rates was noticed. This flattened the curve and kept the number of daily new infections low, as shown in \cite{NYCgithub}. 

The rates of hospitalizations and deaths were smaller by the end of August than in earlier periods of the Covid-19 outbreak in NYC. This phenomenon may be caused by changes in disease virulence or in the protocols used to address Covid-19, like smaller onset to hospitalization meantime, more precise Covid-19 diagnosis, or the introduction of more effective treatments \cite{whoCortico}. Moreover, in NYC, the number of Covid-19 daily tests has been
increasing consistently since the beginning of the outbreak in the end of February-2020. To account for such changes in the aforementioned rates, the corresponding model parameters were adjusted using ratios of reported data, incorporating this feature. This increased considerably the accuracy of the model predictions of hospitalizations and deaths, as the results show (Fig. \ref{fig:NYC1} and Tab.~\ref{tab:NYC1}).

Concerning reopening strategies, some simulated scenarios generated with calibrated parameters indicate that there is no completely safe way of reopening schools, boroughs, or the entire city unless people respect strict protocols of social distance, avoiding direct personal contact. As model predictions show, even when only a borough or only schools are reopened, new infection waves may occur, forcing public authorities to establish contention measures again. While infective individuals are present in a population, there is the risk of new infection waves in reopening strategies, since it is not possible to guarantee that everyone will respect the protocols of contention. Such phenomenon was recently seen in the news \cite{NYT,euronews}. Thus, reopening must be undertaken with strict control of disease transmission, while applying massive testing and enforcing social distance measures. 

Despite the fact that the role of children and teenagers in Covid-19 spread is still unknown, some recent events of disease spread amongst youth population in an overnight camp in Georgia \cite{CDC2020} and in schools in Israel \cite{NYT} indicate that, although most individuals present mild symptoms, they can infect other people. So, reopening schools also represents a risk of new infection waves.

Even New Zealand and China, that successfully contained Covid-19 spread and had long periods without registering community transmission, are now facing new
cases \cite{bbc2020,ft2020}. All these news corroborate our model predictions, suggesting that, without strict control of Covid-19 through social distance or after massive and effective vaccination, a completely safe reopening may be impossible.

It is worth mentioning that, the present model is also suited to simulate and analyze vaccination strategies appropriately, since it addresses dependency on age and spatial distribution. This possibility shall be addressed in a forthcoming article.

\subsection*{Tracking the Reopening of NYC}
On May 2020, the State of New York initiated a four-phase reopening program. NYC joined the fourth phase on 20-July-2020. By 22-Aug-2020, schools and shopping malls were still closed, yet public transportation and a number of economic activities were already operational \cite{NYTReopening,NYFWD}. Strict social distancing measures were still enforced and public authorities were following closely their observance. This kept the number of daily new infections stable. 

Figure~\ref{fig:NYC1} shows the Covid-19 situation in NYC until 21-Aug-2020. The panel presents the comparison of model predictions and reports of daily new infections, as well as the time-dependent basic reproduction rate $R(t)$. Observe that since 19-Mar-2020, $R(t)$ stays around one, meaning the disease transmission is under control, but not eradicated, and that new infection waves may still occur. Even after reaching the fourth phase of controlled reopening, NYC authorities managed to keep transmission under control through social distancing measures, by limiting the operational capacity of numerous services, disinfecting public transportation, and many other practices. 

Notice that, even if Covid-19 is eradicated in NYC, while no effective vaccine is ready to be used in the entire globe, contention measures inside NYC must be kept to avoid new outbreaks caused by the reintroduction of the disease from abroad.

\section{Concluding Remarks}

SEIR-type models have been proposed by a number of authors to predict qualitative aspects of the dynamics of infectious diseases in general, and of the Covid-19 pandemic in particular. See \cite{Bertozzi16732} for a recent account of the SIR models and its connections to other models. Yet, to address the elusive aspects of the complex human interactions within the terrain, we 
feel that one has to forego parsimonious models. 
Functional and high-dimensional models have been used in a number of areas ranging from Financial Mathematics to Population Dynamics \cite{AlbZub2018,pert2007}. They are directly connected with the mathematical theory of Inverse Problems \cite{somersalo,calvetti2020,somersalo2020,ern}.
The present article explores several aspects of infectious diseases, including Covid-19, that have been receiving little attention in the recent literature. We consider time-dependent rates of transmission, hospitalization, and deaths,  as well as the disease age- and gender-dependent severity and transmission, while taking in account the spatial distribution of population.

The model was extensively tested with real data from NYC and its boroughs. 
After calibration, it matched the curves of daily new infections and provided accurate predictions for the number of daily hospitalizations and deaths. It also properly detected the change in the transmission pattern on 19-Mar-2020 caused by contention measures taken on
12-Mar-2020. Moreover, it illustrated the stabilization of the time-dependent basic reproduction rate around the value $1$ in NYC.

Concerning prediction of new infections, the model was also evaluated while using real data and calibrated parameters. It generated
accurate results under controlled and uncontrolled transmission contexts. Moreover, different scenarios as reopening of schools and of an entire borough of NYC were illustrated. In both cases, transmission rates increased considerably, demanding new contention measures. In other words, without reaching herd immunity or the complete disease eradication, we can always face the risk of new infection waves.

Our proposed model is sufficiently general to track transmission dynamics with dependence on age range, gender, and spatial distribution, while evaluating the disease impact on the population. Thus, it is a powerful tool to evaluate scenarios and to build proper vaccination strategies.

\section*{Acknowledgments} JPZ thanks Khalifa University, the Government of Abu Dhabi, and {\it Funda\c{c}\~ao Carlos Chagas Filho de Amparo \`a Pesquisa do Estado do Rio de Janeiro} (FAPERJ) through the program {\it Cientistas do Nosso Estado} for the support during the course of this research.
 JPZ would like to acknowledge very fruitful discussions with Profs.  Dimitris Goussis and Leontios Hadjileontiadis (Khalifa University).  

\bibliographystyle{amsplain}

\providecommand{\bysame}{\leavevmode\hbox to3em{\hrulefill}\thinspace}
\providecommand{\MR}{\relax\ifhmode\unskip\space\fi MR }
\providecommand{\MRhref}[2]{%
  \href{http://www.ams.org/mathscinet-getitem?mr=#1}{#2}
}
\providecommand{\href}[2]{#2}
\begin{thebibliography}{10}

\bibitem{AlbZub2018}
V.~Albani and J.~Zubelli, \emph{{A Splitting Strategy for the Calibration of
  Jump-Diffusion Models}}, Finance and Stochastics \textbf{24} (2020),
  677--722.

\bibitem{tsakris2020}
Cleo Anastassopoulou, Lucia Russo, Athanasios Tsakris, and Constantinos
  Siettos, \emph{{Data-based analysis, modelling and forecasting of the
  COVID-19 outbreak}}, PloS one \textbf{15} (2020), no.~3, e0230405.

\bibitem{bbc2020}
BBC, \emph{{New Zealand coronavirus: 14 new Covid-19 cases reported}}, August
  2020, - \url{https://www.bbc.com/news/world-asia-53761122}.

\bibitem{Bertozzi16732}
Andrea~L. Bertozzi, Elisa Franco, George Mohler, Martin~B. Short, and Daniel
  Sledge, \emph{The challenges of modeling and forecasting the spread of
  covid-19}, Proceedings of the National Academy of Sciences \textbf{117}
  (2020), no.~29, 16732--16738.

\bibitem{gender1}
Sunil~S. Bhopal and Raj Bhopal, \emph{{Sex differential in COVID-19 mortality
  varies markedly by age}}, The Lancet \textbf{396} (2020), no.~10250,
  532--533.

\bibitem{somersalo2020}
Daniela Calvetti, Alexander Hoover, Johnie Rose, and Erkki Somersalo,
  \emph{{Bayesian dynamical estimation of the parameters of an SE(A)IR COVID-19
  spread model}}, https://arxiv.org/abs/2005.04365, 2020.

\bibitem{calvetti2020}
Daniela Calvetti, Alexander~P. Hoover, Johnie Rose, and Erkki Somersalo,
  \emph{{Metapopulation Network Models for Understanding, Predicting, and
  Managing the Coronavirus Disease COVID-19}}, Frontiers in Physics \textbf{8}
  (2020), 261.

\bibitem{covid2020severe}
COVID-19~Response~Team CDC, \emph{{Severe outcomes among patients with
  coronavirus disease 2019 (COVID-19)—United States, February 12--March 16,
  2020}}, MMWR Morb Mortal Wkly Rep \textbf{69} (2020), no.~12, 343--346.

\bibitem{imfTrack}
S.~Chen, D.~Igan, N.~Pierri, and A.~F. Presbitero, \emph{{Tracking the Economic
  Impact of COVID-19 and Mitigation Policies in Europe and the United States}},
  Tech. Report WP/20/125, International Monetary Fund (IMF Working Paper), July
  2020.

\bibitem{chowell2017}
Gerardo Chowell, \emph{Fitting dynamic models to epidemic outbreaks with
  quantified uncertainty: a primer for parameter uncertainty, identifiability,
  and forecasts}, Infectious Disease Modelling \textbf{2} (2017), no.~3,
  379--398.

\bibitem{NYCgithub}
New~York City, \emph{{COVID-19: Data}}, 2020, -
  \url{https://www1.nyc.gov/site/doh/covid/covid-19-data.page}.

\bibitem{BaruchColege}
Baruch College, \emph{{NYC Data}}, 2020, -
  \url{https://www.baruch.cuny.edu/nycdata/population-geography/age_distribution.htm}.

\bibitem{dehning2020}
Jonas Dehning, Johannes Zierenberg, F.~Paul Spitzner, Michael Wibral,
  Joao~Pinheiro Neto, Michael Wilczek, and Viola Priesemann, \emph{{Inferring
  change points in the spread of COVID-19 reveals the effectiveness of
  interventions}}, Science \textbf{369} (2020), no.~6500.

\bibitem{Diekmann1990}
O.~Diekmann, J.~A.~P. Heesterbeek, and J.~A.~J. Metz, \emph{On the definition
  and the computation of the basic reproduction ratio ${R}_0$ in models for
  infectious- diseases in heterogeneous populations}, Journal of Mathematical
  Biology \textbf{28} (1990), no.~4, 365--382.

\bibitem{ern}
H.~Engl, M.~Hanke, and A.~Neubauer, \emph{{Regularization of {I}nverse
  {P}roblems}}, {Mathematics and its Applications}, vol. 375, Kluwer Academic
  Publishers Group, Dordrecht, 1996.

\bibitem{euronews}
Euronews, \emph{Europe's second wave? country-by-country breakdown of resurging
  covid-19 cases}, 2020, -
  \url{https://www.euronews.com/2020/08/06/is-europe-having-a-covid-19-second-wave-country-by-country-breakdown}.

\bibitem{gatto2020}
Marino Gatto, Enrico Bertuzzo, Lorenzo Mari, Stefano Miccoli, Luca Carraro,
  Renato Casagrandi, and Andrea Rinaldo, \emph{Spread and dynamics of the
  covid-19 epidemic in italy: Effects of emergency containment measures},
  Proceedings of the National Academy of Sciences \textbf{117} (2020), no.~19,
  10484--10491.

\bibitem{NYTReopening}
M.~Gold and M.~Stevens, \emph{{What Restrictions on Reopening Remain in New
  York?}}, The New York Times, August 2020.

\bibitem{grasselli2020}
Giacomo Grasselli, Alberto Zangrillo, Alberto Zanella, Massimo Antonelli, Luca
  Cabrini, Antonio Castelli, Danilo Cereda, Antonio Coluccello, Giuseppe Foti,
  Roberto Fumagalli, et~al., \emph{{Baseline characteristics and outcomes of
  1591 patients infected with SARS-CoV-2 admitted to ICUs of the Lombardy
  region, Italy}}, JAMA \textbf{323} (2020), no.~16, 1574--1581.

\bibitem{guan2020}
Wei-jie Guan, Zheng-yi Ni, Yu~Hu, Wen-hua Liang, Chun-quan Ou, Jian-xing He,
  Lei Liu, Hong Shan, Chun-liang Lei, David~SC Hui, et~al., \emph{{Clinical
  characteristics of coronavirus disease 2019 in China}}, New England journal
  of medicine \textbf{382} (2020), no.~18, 1708--1720.

\bibitem{huang2020}
Chaolin Huang, Yeming Wang, Xingwang Li, Lili Ren, Jianping Zhao, Yi~Hu,
  Li~Zhang, Guohui Fan, Jiuyang Xu, Xiaoying Gu, et~al., \emph{{Clinical
  features of patients infected with 2019 novel coronavirus in Wuhan, China}},
  The Lancet \textbf{395} (2020), no.~10223, 497--506.

\bibitem{gender2}
Jian-Min Jin, Peng Bai, Wei He, Fei Wu, Xiao-Fang Liu, De-Min Han, Shi Liu, and
  Jin-Kui Yang, \emph{{Gender Differences in Patients with COVID-19: Focus on
  Severity and Mortality}}, Frontiers in Public Health \textbf{\;} (2020),
  no.~\;, \;.

\bibitem{keeling2008}
M.J. Keeling and R.~Rohani, \emph{Modeling infectious diseases in humans and
  animals}, Princeton University Press, 2008.

\bibitem{KerMack1927}
William~Ogilvy Kermack and Anderson~G McKendrick, \emph{A contribution to the
  mathematical theory of epidemics}, Proc. R. Soc. Lond. A \textbf{115} (1927),
  no.~772, 700--721.

\bibitem{NYT}
I.~Kershner and P.~Belluck, \emph{When covid subsided, israel reopened its
  schools. it didn’t go well.}, The New York Times, 2020.

\bibitem{Incubation1}
Stephen~A. Lauer, Kyra~H. Grantz, Qifang Bi, Forrest~K. Jones, Qulu Zheng,
  Hannah~R. Meredith, Andrew~S. Azman, Nicholas~G. Reich, and Justin Lessler,
  \emph{{The Incubation Period of Coronavirus Disease 2019 (COVID-19) From
  Publicly Reported Confirmed Cases: Estimation and Application}}, Annals of
  Internal Medicine \textbf{172} (2020), no.~9, 577--583.

\bibitem{pert2007}
Beno{\^i}t Perthame, \emph{{Transport {E}quations in Biology}}, Birkh{\"a}user
  Verlag, 2007.

\bibitem{ft2020}
C.~Shepherd, \emph{China records biggest one-day rise in coronavirus cases
  since march}, Financial Times, 2020.

\bibitem{somersalo}
Erkki Somersalo and Jari Kapio, \emph{{{S}tatistical and {C}omputational
  {I}nverse {P}roblems}}, {Applied Mathematical Sciences}, vol. 160, Springer,
  2004.

\bibitem{NYFWD}
New~York State, \emph{{New York Forward}}, 2020, -
  \url{https://forward.ny.gov/}.

\bibitem{CDC2020}
C.M. Szablewski, K.T. Chang, M.M. Brown, and et~al., \emph{{SARS-CoV-2
  Transmission and Infection Among Attendees of an Overnight Camp -- Georgia,
  June 2020}}, MMWR Morb Mortal Wkly Rep \textbf{69} (2020), 1023--1025.

\bibitem{verity2020}
Robert Verity, Lucy~C Okell, Ilaria Dorigatti, Peter Winskill, Charles
  Whittaker, Natsuko Imai, Gina Cuomo-Dannenburg, Hayley Thompson, Patrick~GT
  Walker, Han Fu, et~al., \emph{Estimates of the severity of coronavirus
  disease 2019: a model-based analysis}, The Lancet Infectious Diseases
  \textbf{20} (2020), no.~6, 669--677.

\bibitem{wef}
WEF, \emph{{IMF: New predictions suggest a deeper recession and a slower
  recovery}}, 2020, -
  \url{https://www.weforum.org/agenda/2020/06/imf-lockdown-recession-covid19-coronavirus-economics-recession/}.

\bibitem{whoCortico}
WHO, \emph{{Corticosteroids for COVID-19}}, Tech. report, World Heath
  Organization, September 2020.

\bibitem{who2020}
\bysame, \emph{{Report of the WHO-China joint mission on coronavirus disease
  2019 (COVID-19)}}, 2020.

\bibitem{wu2020}
Joseph~T Wu, Kathy Leung, Mary Bushman, Nishant Kishore, Rene Niehus, Pablo~M
  de~Salazar, Benjamin~J Cowling, Marc Lipsitch, and Gabriel~M Leung,
  \emph{{Estimating clinical severity of {COVID}-19 from the transmission
  dynamics in Wuhan, China}}, Nature Medicine \textbf{26} (2020), no.~4,
  506--510.

\bibitem{wu2020c}
Joseph~T Wu, Kathy Leung, and Gabriel~M Leung, \emph{Nowcasting and forecasting
  the potential domestic and international spread of the 2019-ncov outbreak
  originating in wuhan, china: a modelling study}, The Lancet \textbf{395}
  (2020), no.~10225, 689--697.

\bibitem{wu2020b}
Zunyou Wu and Jennifer~M McGoogan, \emph{Characteristics of and important
  lessons from the coronavirus disease 2019 (covid-19) outbreak in china:
  summary of a report of 72 314 cases from the chinese center for disease
  control and prevention}, JAMA \textbf{323} (2020), no.~13, 1239--1242.

\end{thebibliography}
\providecommand{\bysame}{\leavevmode\hbox to3em{\hrulefill}\thinspace}
\providecommand{\MR}{\relax\ifhmode\unskip\space\fi MR }
\providecommand{\MRhref}[2]{%
  \href{http://www.ams.org/mathscinet-getitem?mr=#1}{#2}
}
\providecommand{\href}[2]{#2}

\end{document}